# A Novel Interactive Two-stage Joint Retail Electricity Market for Multiple Microgrids

Chunyi Huang, *Member, IEEE*, Mingzhi Zhang, *Graduate Student Member, IEEE,* Chengmin Wang, Ning Xie, Jian Wang, Shi Peng

***Abstract*—To accommodate the advent of microgrids (MG) managing distributed energy resources (DER) in distribution systems, an interactive two-stage joint retail electricity market mechanism is proposed to provide an effective platform for these prosumers to proactively join in retail transactions. Day-ahead stochastic energy trading between the distribution system operator (DSO) and MGs is conducted in the first stage of a centralized retail market, where a chance-constrained uncertainty distribution locational marginal price (CC-UDLMP) containing the cost of uncertainty precautions is used to settle transactions. In the second stage, a novel intra-day peer-to-peer-based (P2P) flexibility transaction pattern is implemented between MGs in local flexibility markets under the regulation of DSO to eliminate power imbalances caused by rolling-based estimates whilst considering systematic operations. A fully distributed iterative algorithm is presented to find the equilibrium solution of this two-stage sequential game framework. Moreover, in order to enhance the versatility of this algorithm, an improved Lp-box alternating direction methods of multipliers (ADMM) algorithm is used to efficiently resolve the first-stage stochastic economic dispatch problem with a mixed-integer second-order cone structure. It is verified that the proposed market mechanism can effectively improve the overall market efficiency under uncertainties.**

## NOMENCLATURE

| | |
|---|---|
| *A. Indices* | |
| $u$ | Uncertainty index. |
| $t$ | Time index. |
| *B. Sets* | |
| $\Psi_u$ | Set of uncertainty in network and MGs. |
| $\mathrm{T}$ | Number of time slots over time horizon. |
| *C. Parameters* | |
| $\mathrm{c}_{1\mathrm{d}}, \mathrm{c}_{2\mathrm{d}}$ | Two-stage dispatching costs of networked DERs and upstream power injection. |
| $\mathrm{c}_{\mathrm{nl}}$ | Equivalent network loss cost. |
| $\mathrm{c}_{\mathrm{ppc}}$ | Flexibility adjustment costs at PCC from DSO. |
| $\mathrm{c}_{1\mathrm{m}}, \mathrm{c}_{2\mathrm{m}}$ | Two-stage dispatching costs of DERs within MGs. |
| $\mathrm{c}_{1\mathrm{pc}}, \mathrm{c}_{2\mathrm{pc}}$ | Two-stage load compensations of users in MGs. |
| $\mathrm{c}_{\mathrm{fp}}$ | Procurement costs of P2P flexibility contracts. |
| $\mathrm{c}_{\mathrm{fs}}$ | Flexibility adjustment costs required by MGs. |
| $\overline{\mathrm{P}}^{\mathrm{LM}}, \overline{\mathrm{Q}}^{\mathrm{LM}}$ | Deterministic load demand of MGs. |
| $\zeta$ | Acceptable rate of load shedding. |
| $\tan\theta$ | Power factor of load demand in MGs. |
| $\alpha$ | Penalty coefficient in Lp-box ADMM algorithm. |
| $i_{pe}$ | Number of iterations with constant penalty terms. |
| *D. Variables* | |
| $\tilde{P}^{\mathrm{G}}, \check{P}^{\mathrm{G}}$ | Stochastic available active output of DERs. |
| $\tilde{P}^{\mathrm{L}}, \check{P}^{\mathrm{L}}, \check{\mathrm{P}}^{\mathrm{LM}}$ | Stochastic load demands of network and MGs. |
| $x^{\mathrm{pg}}, x^{\mathrm{qg}}$ | Day-ahead outputs of networked DERs and upstream power injection. |
| $x^{\mathrm{s}}$ | Day-ahead state of charge of networked ESS. |
| $\tilde{x}^{\mathrm{df}}$ | Probabilistic systematic variables in undirected SOC-based distribution AC power flow model. |
| $z^{+}, z^{-}$ | Binary power direction variables. |
| $z^{\mathrm{ch}}, z^{\mathrm{dic}}$ | Binary operating status of networked ESS. |
| $\overline{x}^{\mathrm{pg}}, \underline{x}^{\mathrm{pg}}$ | Upward- and downward- reserve capacity of DSO. |
| $\hat{x}^{\mathrm{pg+}}, \hat{x}^{\mathrm{pg-}}$ | Upward- and downward- reserve output of DSO. |
| $\hat{x}^{\mathrm{qg}}$ | Reactive adjustment of networked DERs. |
| $\hat{x}^{\mathrm{s}}$ | Intra-day state of charge of networked ESS. |
| $\hat{x}^{\mathrm{ppc+}}, \hat{x}^{\mathrm{ppc-}}$ | Flexibility adjustments at PCC from DSO. |
| $\hat{x}^{\mathrm{qpc}}$ | Intra-day reactive adjustments at PCC from DSO. |
| $\hat{x}^{\mathrm{df}}$ | Intra-day systematic variables of SOC model. |
| $y^{\mathrm{ppc}}, y^{\mathrm{qpc}}$ | Day-ahead import power at PCC from DSO. |
| $y^{\mathrm{pg}}, y^{\mathrm{qg}}$ | Day-ahead outputs of DERs in MGs. |
| $y^{\mathrm{s}}$ | Day-ahead state of charge of ESS in MGs. |
| $y^{\mathrm{Lp}}, y^{\mathrm{Lq}}$ | Day-ahead load shedding of MGs. |
| $y^{\mathrm{zch}}, y^{\mathrm{zdic}}$ | Binary operating status of ESS in MGs. |
| $\overline{y}^{\mathrm{pg}}, \underline{y}^{\mathrm{pg}}$ | Upward- and downward- reserve capacity of MGs. |
| $\hat{y}^{\mathrm{pf}}, \hat{y}^{\mathrm{qf}}$ | P2P flexibility services between MGs. |
| $\hat{y}^{\mathrm{pg+}}, \hat{y}^{\mathrm{pg-}}$ | Upward- and downward- reserve output of MGs. |
| $\hat{y}^{\mathrm{qg}}$ | Reactive adjustment output of DERs in MGs. |
| $\hat{y}^{\mathrm{ppc+}}, \hat{y}^{\mathrm{ppc-}}$ | Flexibility adjustments at PCC from MGs. |
| $\hat{y}^{\mathrm{qpc}}$ | Intra-day reactive adjustments at PCC from MGs. |
| $\hat{y}^{\mathrm{s}}$ | Intra-day state of charge of ESS in MGs. |
| $\hat{y}^{\mathrm{Lp+}}, \hat{y}^{\mathrm{Lp-}}$ | Intra-day adjustment load shedding of MGs. |
| $\hat{y}^{\mathrm{Lq}}$ | Intra-day reactive load shedding of MGs. |

## I. INTRODUCTION

THE integration of DERs promotes the transition of the global energy consumption structure. Due to the lack of advanced metering and controlling systems, current practices for DER management are mainly based on centralized control and uniform subsidies. The deregulation of retail markets breaks the vertical monopoly of distribution systems, enabling DER entities to proactively join in markets. Accordingly, MGs that prioritize internal autonomous operations and also arbitrage in energy trading are promising prosumers to manage large shares of DERs with higher economic efficiency [1]-[2].

Existing studies have explored the market mechanism of MG as a price-maker [3]. To improve the reliability of power supply, a hierarchical model is established in [4]-[5] to coordinate the complementarity of MG clusters. In the energy management market strategy for islanded MGs in [6], a local demand center is founded to distribute exchanged power so as to increase the overall revenue. In addition, a tri-settlement incentive market mechanism for multiple MGs is presented in [7], which deconstructs power delivery into fragmented deterministic transactions and introduces demand response to hedge the uncertainty risk by employing Nash bargaining among MGs.

The work on the MG cluster business model has pervasively neglected the network topology and system-level operation restrictions, undoubtedly exposing two practical drawbacks in the MG management. On the one hand, with the explosive growth of DERs in MGs, the energy sharing scheme within MG clusters in this mode may trigger security issues in distribution systems. On the other hand, previous studies of MG clusters have placed holistic stability over market profitability, which ignores differentiated decision preferences of MGs, inevitably leading to losses of individual benefits. Hence, a novel retail market mechanism is needed to strengthen interactions between DSO and various MGs in order to improve market efficiency and ensure systematic security under uncertainty.

A sensitive distribution locational marginal price (DLMP) reflecting resource scarcity and systemic operating status costs [8] is necessitated to incentivize timely quotes of MGs tracking peak-shaving regulations of the distribution side. Limited by the nonconvexity of AC power flow, existing DLMP formulations are divided into linearization-based [9] and second-order cone (SOC) based methods [10]. The cone-based method outperforms the former in computational performance, whereas its accuracy is assured for radial distribution networks only under a set of sufficient conditions [11]-[12]. To overcome inherent deficiencies of the classic SOC-based model in describing uncertain distribution power flow, an undirected SOC-based distribution AC power flow formulation is proposed in our previous work [13], which introduces binary variable pairs to discretize possible power flow solutions and thus removing a priori requirement for power flow paths. Apart from deterministic DLMP calculation based on scenario-based uncertainty modeling, some recent studies have applied uncertainty DLMP (UDLMP) to further assess the impact of uncertainty on market settlements. By leveraging stochastic models to characterize renewable energy generations and load demands, a chance-constrained UDLMP formulation is proposed in [14] to calculate energy and balancing regulation costs considering supply uncertainties and risk tolerances. Moreover, a robust model is used in [15] to formulate a two-step congestion management method for flexible buildings. Nevertheless, existing UDLMP formulations neither analyze the extensive uncertainty nor adequately examine the impact of nodal disturbances on systematic operating conditions, indicating the gap in the study of generic UDLMP stochastic calculations.

Besides, to further handle the unpredictability within the distribution network and MGs, the balancing capacity sharing mechanism is regarded as a cost-effective alternative to reduce the amount of network-side reserves required under the centralized regulation mode. As a branch of transactive energy [16], P2P-based transactions advocate profitable decisions of individual entities, which encompass the coordinated [17], the community [18], and the decentralized frameworks [19] following a central to decentralized communication pattern. To support diverse commercial practices while maintaining the supervision of DSO, the local flexibility market (LFM) was proposed in recent initiatives, serving as a platform for local participants to trade flexibility services [20]-[21]. Despite these models have explored possible flexibility sharing among entities, few examined the combined application of P2P trading and LFM in flexibility support between DSO and MGs, while fully exploiting the flexibility potential of the end side.

To address abovementioned gaps of existing studies, an interactive two-stage joint retail market mechanism for multiple MGs is presented in this work. A two-stage sequential game that enables day-ahead and intra-day bargains between DSO and MGs is built to reconcile systematic security considerations and natural arbitrages of MGs. In the first stage, the day-ahead energy transactions are implemented in a centralized retail market in face of stochastic uncertainties, where the CC-UDLMP is adopted to guide strategic behaviors of MGs. The proposed CC-UDLMP overcomes the limitation of current uncertainty DLMP calculations by quantifying the impact of probabilistic supply and demand uncertainties and system status on market-clearing. In order to eliminate power imbalances caused by intra-day rolling forecasts, in the second stage, flexibility trading among MGs is conducted in several LFMs under the supervision of DSO. This novel business model coordinates the P2P transactions between MGs and systematic operating calibration executed by DSO. At last, as this incentive framework is computationally intractable, a fully distributed iterative algorithm is developed to address this problem with privacy concerns. In summary, the main contributions of this paper are threefold.

① An interactive two-stage joint retail market that integrates centralized and local trading is constructed, with multiple MGs submitting strategic biddings for energy and flexibility services under a certain supervision of DSO.

② A chance-constrained uncertainty distribution locational marginal pricing method and its equivalent transformation is proposed, where the cost of systematic preventive measures for extensive uncertainties is internalized in the settlement.

③ A fully distributed iterative algorithm is proposed to address this two-stage sequential game model, where an improved Lp-box ADMM distributed algorithm is applied to enhance the versatility of this work.

The rest of this article is organized as follows. Section II elaborates the two-stage stochastic uncertainty modeling and the retail market mechanism. Section III presents mathematical formulations of this market framework. Section IV illustrates the CC-UDLMP and its transformations. Section V provides a fully distributed iterative solving algorithm for this two-stage retail market along with the convergence analysis. Section VI provides informative case studies, while the conclusion and prospective works are presented in Section VII.

## II. Two-stage Joint Retail Electricity Market

### A. Two-stage Modeling of Uncertainty

The uncertainty $u \in \Psi_u$ associated with daily available outputs of wind and photovoltaic (PV) DERs along with nodal load demands is uniformly described by a two-stage stochastic model. It is first modeled as a known probability distribution and then estimated on a rolling basis, depicted in Fig. 1.

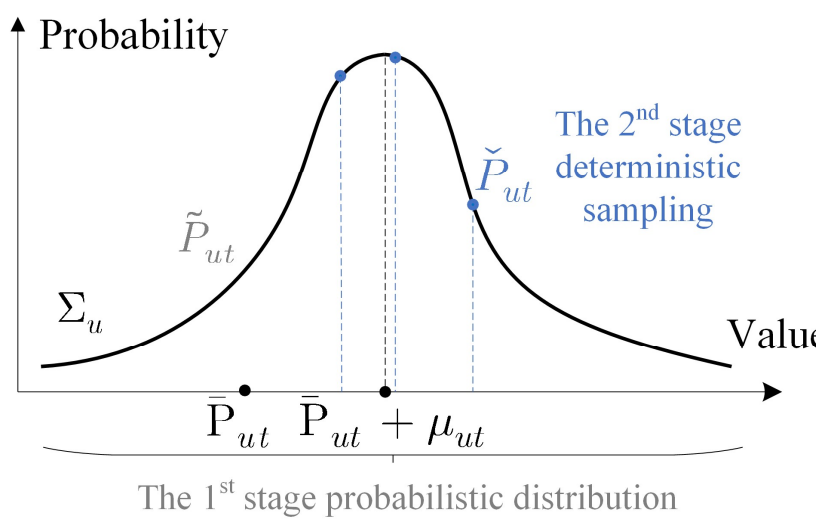

Fig. 1. Two-stage stochastic modeling of uncertainties

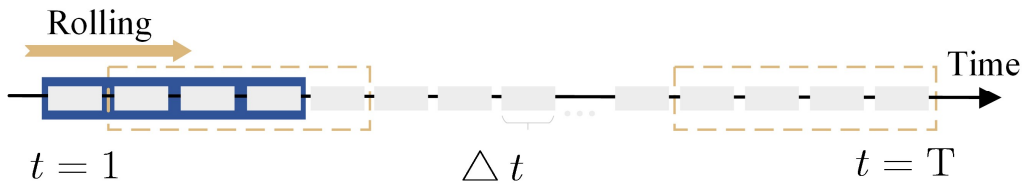

Fig. 2. Second-stage rolling estimation of uncertainties

As shown in (1), the hourly uncertainty $\tilde{P}_{ut}$ is composed of a typical value $\bar{\mathrm{P}}_{ut}$ and a random deviation $\hat{P}_{ut}$. Given temporal correlations of uncertainty, each $\hat{P}_{ut}$ is assumed to obey a multivariate Gaussian distribution with mean $\mu_{ut}$ and covariance matrix $\Sigma_u$ related to $\hat{P}_{ut}$ over the time horizon. With the fixed power factor assumption, the nodal reactive load demand is formed as (2), while the reactive power availability of DER is individually controlled and is not considered as an uncertainty.

$$\tilde{P}_{ut} = \bar{\mathrm{P}}_{ut} + \hat{P}_{ut}, \hat{P}_{ut} \propto (\mu_{ut}, \Sigma_u) \tag{1}$$

$$\tilde{Q}_{ut} = \tilde{P}_{ut}\tan\theta, \quad \forall u \in \Psi_u, t \in \mathrm{T} \tag{2}$$

In the second stage, the uncertainties will be further corrected in a rolling pattern a few hours before it occurs. As depicted in Fig. 2, the size of sliding window is set as 4h and the step size is 1h. It is clear that the rolling estimation of $\check{P}_{ut}, \check{Q}_{ut}$ will be randomly located within the distribution of Fig.1.

### *B. Interactive Two-stage Retail Electricity Market mechanism*

The current retail market typically relies on top-down management by DSO to dispatch network-side reserves so as to maintain real-time system security. This exclusive regulation approach ignores the adaptive elasticity of prosumers in tracking dynamic tariffs and fails to take advantage of the flexibility potential of end-use entities. To address these bottlenecks, we propose a two-stage joint retail market that integrates centralized and local trading among DSO and MGs to accommodate uncertainties. It is noted that the DERs therein include distributed wind and PV units, small-scale energy storage systems (ESS), and reactive power compensators.

As shown in Fig.3, the first-stage day-ahead energy trading between DSO and individual MGs is conducted in the centralized retail market, using CC-UDLMP for settlement. At the second stage, intra-day flexibility contracts between MGs are formed by P2P sharing in LFMs under the supervision of DSO, where local flexibility resources are deployed to reduce reliance on pre-arranged networked reserves. This hierarchical framework can be viewed as a two-stage sequential game, where DSO develops dispatch schemes and broadcasts dynamic tariffs, and then multiple MGs report their outputs at the point of common coupling (PCC) and also formulate energy and flexibility contracts that evoke decision revisions from DSO.

Specifically, in the first stage, DSO whose goal is to minimize procurement costs under uncertainty probabilistic distributions, will initially develop a stochastic day-ahead dispatch scheme and broadcast the CC-UDLMP based on estimated demands of MGs. Hereafter, MGs will perform economic scheduling of internally managed assets and submit aggregated demand curves at PCC to DSO for arbitrage in the currently disclosed settlement. This process will be repeated until the first-stage Nash equilibrium including dispatch and reserve capacity plans is attained. The subsequent second stage is intra-day flexibility transactions to dissolve the imbalanced power caused by rolling estimates of uncertainty. Flexibility services are divided into upward and downward services, where the upward flexibility is traded between MGs to address unexpected power supply shortages. At first, MG operators will negotiate with each other at LFMs to form P2P flexibility contracts based on reserve capacity retained previously. Then, upon the receipt of requested flexibility proposals, DSO will re-dispatch networked reserves based on updated rolling estimates and formulate flexibility exchange schemes with MGs to ensure that the execution of P2P transactions will not jeopardize the stability of the overall system. After limited interactions, a final second-stage settlement scheme is obtained.

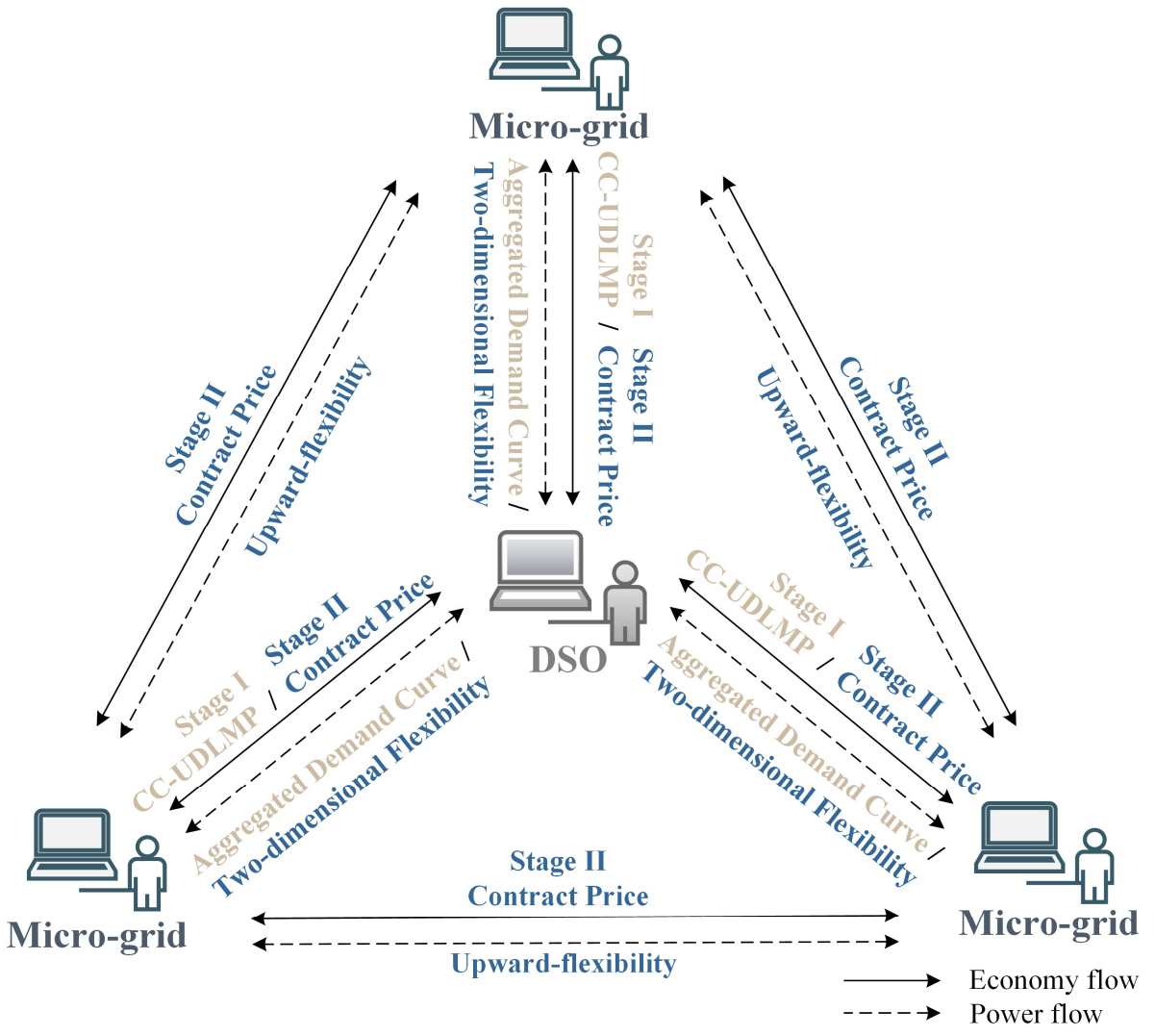

Fig. 3. Framework of two-stage joint retail electricity market

## III. Mathematical Formulation of Iterative Two-stage Joint Retail Electricity Market

### *A. Two-stage Market Settlement Model of DSO*

The two-stage stochastic market settlement of DSO is implemented in succession to clear system-level energy and flexibility services. Detailed formulations of this settlement model are provided in Appendix A.

In the first stage, a chance-constrained stochastic economic dispatch model is built in the face of uncertainty probabilistic distributions detailed as $\tilde{P}^{\mathrm{G}}, \tilde{P}^{\mathrm{L}}, \tilde{Q}^{\mathrm{L}}$, while the compact-form formulation is listed in (3)-(12). The bold letters such as $\mathbf{A_d}$ are coefficient matrixes or column vectors, I stands for unit matrix, and superscript $T$ represents the transposition of the matrix.

$$\min_x \mathrm{c}_{1\mathrm{d}}^T x^{\mathrm{pg}} + \mathrm{c}_{\mathrm{nl}}^T \tilde{x}^{\mathrm{df}} \tag{3}$$

*subject to.*

$$(x, z) \in \{x^{\mathrm{pg}}, x^{\mathrm{qg}}, x^{\mathrm{s}}, \tilde{x}^{\mathrm{df}}, z^{+}, z^{-}, z^{\mathrm{ch}}, z^{\mathrm{dic}}, \overline{x}^{\mathrm{pg}}, \underline{x}^{\mathrm{pg}}\} \tag{4}$$

$$\mathbf{A_d} x^{\mathrm{pg}} + \mathbf{B_{1d}} \tilde{x}^{\mathrm{df}} - [\![ y^{\mathrm{ppc}} ]\!] = \tilde{P}^{\mathrm{L}} : \tau^{\mathrm{p}}, \tag{5.a}$$

$$\mathbf{C_d}x^{\mathrm{qg}}+\mathbf{D_{1d}}\tilde{x}^{\mathrm{df}}-[\![y^{\mathrm{qpc}}]\!]=\tilde{Q}^{\mathrm{L}}, \tag{5.b}$$

$$\mathbf{E_d}x^{\mathrm{pg}}+\mathbf{F_{1d}}\tilde{x}^{\mathrm{df}}+\mathbf{G}_{\mathrm{d}}x^{\mathrm{s}}=\xi, \tag{6}$$

$$|\mathbf{H_{1d}}\tilde{x}^{\mathrm{df}}|\preccurlyeq 0, \tag{7}$$

$$\mathrm{Prob}\big(-\mathbf{J_d}z^{-}\le\mathbf{K_{1d}}\tilde{x}^{\mathrm{df}}+\mathbf{L_d}x^{\mathrm{pg}}\le\mathbf{M_d}z^{+}+\mathbf{N_d}\tilde{P}^{\mathrm{G}}\big)\ge 1-\gamma, \tag{8}$$

$$0\le\mathbf{O_d}(x^{\mathrm{pg}}-\underline{x}^{\mathrm{pg}}), \tag{9.a}$$

$$\mathbf{P_d}(x^{\mathrm{pg}}+\bar{x}^{\mathrm{pg}})\le\mathbf{Q_d}z^{\mathrm{ch}}+\mathbf{R_d}z^{\mathrm{dic}}+\mu, \tag{9.b}$$

$$\delta\le\mathbf{S_d}x^{\mathrm{qg}}+\mathbf{U_d}x^{\mathrm{s}}\le\psi, \tag{10}$$

$$z^{-}+z^{+}=\mathrm{I},z^{-},z^{+}\in\{0,1\}, \tag{11}$$

$$z^{\mathrm{ch}}+z^{\mathrm{dic}}\le\mathrm{I},z^{\mathrm{ch}},z^{\mathrm{dic}}\in\{0,1\}. \tag{12}$$

The day-ahead dispatch and reserve allocation schemes are generated in (4) by minimizing the operating costs of network-side generations and losses in (3). It is noted that the interchange power at PCC $[\![y^{\mathrm{ppc}}]\!],[\![y^{\mathrm{qpc}}]\!]$ in (5) is updated each time based on the latest feedback from MGs. Given the variability of AC power flow directions during the bidding process, an undirected second-order cone-based power flow formulation [13] is used in (5)-(6) to overcome the mandatory condition of the classical SOC-based model [22] in terms of predetermining the flow directions. The symbol $\preccurlyeq$ in (7) indicates the rotated second-order cone calculator, while binary directional variables $z^{+},z^{-}$ in (11) are used to limit the uniqueness of the power flow solution. Affected by the uncertainty of supply and demand, the active output of DERs $x^{\mathrm{pg}}$ along with system state variables $\tilde{x}^{\mathrm{df}}$, including the squared nodal voltage amplitude, the squared feeder current, and the active and reactive power flow through feeder are restricted by chance constraints with a certain confidence level in (8). The spinning reserve is supplied by the upstream network injection and ESS in (9). The operational bounding box of other DERs is listed in (10), while the charging and discharging status of ESS are described in (12).

In the following second stage, DSO dispatches retained reserves and trades bi-directional flexibility services with MGs on a rolling basis to ensure network-wide security. The compact model is listed in (13)-(21), and these constraints will be updated with the progressive move of a sliding window.

$$\min_{\hat{x}}\mathrm{c}_{2\mathrm{d}}^{T}(\hat{x}^{\mathrm{pg+}}+\hat{x}^{\mathrm{pg-}})+\mathrm{c}_{\mathrm{ppc}}^{T}(\hat{x}^{\mathrm{ppc+}}+\hat{x}^{\mathrm{ppc-}}) \tag{13}$$

$$\textit{subject to.}\quad \hat{x}\in\{\hat{x}^{\mathrm{pg+}},\hat{x}^{\mathrm{pg-}},\hat{x}^{\mathrm{qg}},\hat{x}^{\mathrm{s}},\hat{x}^{\mathrm{ppc+}},\hat{x}^{\mathrm{ppc-}},\hat{x}^{\mathrm{qpc}},\hat{x}^{\mathrm{df}}\} \tag{14}$$

$$\mathbf{A_d}(x^{\mathrm{pg*}}+\hat{x}^{\mathrm{pg+}}-\hat{x}^{\mathrm{pg-}})-([\![y^{\mathrm{ppc*}}]\!]+\hat{x}^{\mathrm{ppc+}}-\hat{x}^{\mathrm{ppc-}}+[\![y^{\mathrm{pf}}]\!])+\mathbf{B_{2d}}\hat{x}^{\mathrm{df}}=\check{P}^{\mathrm{L}}, \tag{15.a}$$

$$\mathbf{C_d}(x^{\mathrm{qg*}}+\hat{x}^{\mathrm{qg+}})-([\![y^{\mathrm{qpc*}}]\!]+\hat{x}^{\mathrm{qpc}}+[\![y^{\mathrm{qf}}]\!])+\mathbf{D_{2d}}\hat{x}^{\mathrm{df}}=\check{Q}^{\mathrm{L}}, \tag{15.b}$$

$$\mathbf{E_d}(x^{\mathrm{pg*}}+\hat{x}^{\mathrm{pg+}}-\hat{x}^{\mathrm{pg-}})+\mathbf{F_{2d}}\hat{x}^{\mathrm{df}}+\mathbf{G_d}\hat{x}^{\mathrm{s}}=\xi, \tag{16}$$

$$|\mathbf{H_{2d}}\hat{x}^{\mathrm{df}}|\preccurlyeq 0, \tag{17}$$

$$0\le\mathbf{K_{2d}}\hat{x}^{\mathrm{df}}+\mathbf{L_d}(x^{\mathrm{pg*}}+\hat{x}^{\mathrm{pg+}}-\hat{x}^{\mathrm{pg-}})\le\eta+\mathbf{N_d}\check{P}^{\mathrm{G}} \tag{18}$$

$$0\le\hat{x}^{\mathrm{pg+}}\le\overline{x}^{\mathrm{pg}}, \tag{19.a}$$

$$0\le\hat{x}^{\mathrm{pg-}}\le\underline{x}^{\mathrm{pg}}, \tag{19.b}$$

$$\mathbf{O_d}(x^{\mathrm{pg*}}+\hat{x}^{\mathrm{pg+}}-\hat{x}^{\mathrm{pg-}})\le\mathbf{R_d}z^{\mathrm{ch*}}+\mathbf{S_d}z^{\mathrm{dic*}}+\mu, \tag{19.c}$$

$$\delta\le\mathbf{T_d}(x^{\mathrm{qg*}}+\hat{x}^{\mathrm{qg}})+\mathbf{U_d}\hat{x}^{\mathrm{s}}\le\psi, \tag{20}$$

$$-\varepsilon\le[\![y^{\mathrm{ppc*}}]\!]+\hat{x}^{\mathrm{ppc+}}-\hat{x}^{\mathrm{ppc-}}+[\![y^{\mathrm{pf}}]\!]\le\varepsilon, \tag{21.a}$$

$$-\epsilon\le[\![y^{\mathrm{qpc*}}]\!]+\hat{x}^{\mathrm{qpc}}+[\![y^{\mathrm{qf}}]\!]\le\epsilon, \tag{21.b}$$

With the goal of minimizing the operating costs including the cost of spinning reserves and network losses in (13), the reserve scheduling scheme and flexibility exchange plans with MGs are formed in (14). Assuming power flow paths obtained at the first stage will not be altered, the classic SOC-based power model with fixed directions is adopted in (15)-(17), where $\check{P}^{\mathrm{L}},\check{Q}^{\mathrm{L}}$ are rolling estimates of power demand. It is noted that the network-side energy dispatch plan $[\![y^{\mathrm{ppc*}}]\!],[\![y^{\mathrm{qpc*}}]\!]$ and the flexibility regulation of MGs $[\![y^{\mathrm{pf}}]\!],[\![y^{\mathrm{qf}}]\!]$, i.e. the cumulative flexibility regulation at PCC, are both fixed in (15) and (21). With updated availability of clean energy $\check{P}^{\mathrm{G}}$, (18) uniformly limits the operating boundary of DERs and system state variables. (19) restricts the operating range of the reserve within the retained capacity, (20) is the operating condition of other DERs, and (21) is the bounding box limit of total power exchange at PCC.

## *B. Two-stage Bidding of Multiple Microgrids*

MGs accessed at different nodes formulate energy and P2P flexibility biddings sequentially in a two-stage sequential game, whose expanded model is detailed in Appendix B.

In the first phase, each MG will independently develop profitable chance-constrained dispatch schemes to cope with probabilistic uncertainty of internal DER availability. The compact-form model is uniformly elaborated in (22)-(30).

$$\max_{y}\,[\![\tau^{\mathrm{p}}]\!]^{T}(-y^{\mathrm{ppc}})-(\mathrm{c}_{1\mathrm{m}}^{T}y^{\mathrm{pg}}+\mathrm{c}_{1\mathrm{pc}}^{T}y^{\mathrm{Lp}}) \tag{22}$$

$$\textit{subject to.}\, y\in\left\{\begin{matrix}y^{\mathrm{ppc}},y^{\mathrm{qpc}},y^{\mathrm{pg}},y^{\mathrm{qg}},y^{\mathrm{s}},y^{\mathrm{Lp}},y^{\mathrm{Lq}},\\ y^{\mathrm{zch}},y^{\mathrm{zdic}},\overline{y}^{\mathrm{pg}},\underline{y}^{\mathrm{pg}}\end{matrix}\right\} \tag{23}$$

$$\mathbf{A_m}y^{\mathrm{pg}}+y^{\mathrm{ppc}}=\overline{\mathrm{P}}^{\mathrm{LM}}-y^{\mathrm{Lp}}, \tag{24.a}$$

$$\mathbf{C_m}y^{\mathrm{qg}}+y^{\mathrm{qpc}}=\overline{\mathrm{Q}}^{\mathrm{LM}}-y^{\mathrm{Lq}}, \tag{24.b}$$

$$\mathrm{Prob}\big(0\le\mathbf{H_m}y^{\mathrm{pg}}\le\mathbf{J_m}\tilde{P}^{\mathrm{gM}}\big)\ge 1-\gamma, \tag{25}$$

$$\zeta\le\mathbf{D_m}y^{\mathrm{ppc}}+\mathbf{E_m}y^{\mathrm{qpc}}+\mathbf{F_m}y^{\mathrm{qg}}+\mathbf{G_m}y^{\mathrm{s}}\le\lambda, \tag{26}$$

$$0\le\mathbf{K_m}(y^{\mathrm{pg}}-\underline{y}^{\mathrm{pg}}), \tag{27.a}$$

$$\mathbf{L_m}(y^{\mathrm{pg}}+\overline{y}^{\mathrm{pg}})\le\mathbf{M_m}y^{\mathrm{zch}}+\mathbf{N_m}y^{\mathrm{zdic}}+\varpi, \tag{27.b}$$

$$y^{\mathrm{zch}}+y^{\mathrm{zdic}}\le\mathrm{I},y^{\mathrm{zch}},y^{\mathrm{zdic}}\in\{0,1\}, \tag{28}$$

$$\mathbf{P_m}y^{\mathrm{s}}+\mathbf{Q_m}y^{\mathrm{pg}}=\sigma, \tag{29}$$

$$(1-\zeta)\overline{\mathrm{P}}^{\mathrm{LM}}\le\overline{\mathrm{P}}^{\mathrm{LM}}-y^{\mathrm{Lp}}\le\overline{\mathrm{P}}^{\mathrm{LM}}, \tag{30.a}$$

$$y^{\mathrm{Lq}}=y^{\mathrm{Lp}}\tan\theta. \tag{30.b}$$

The day-ahead energy bid of MG is formed in (23) for arbitrage under the time-varying CC-UDLMP published in (22). The revenue in (22) is obtained by subtracting the internal operating costs from the income of energy trading, while the CC-UDLMP at PCC $[\![\tau^{\mathrm{p}}]\!]$ are derived from the dual multiplier of (5.a). The power balance constraint within MG is met by (24). Given the probabilistic availability of renewable energy, the output of DERs is chance-constrained in (25). (26) uniformly limits the available energy exchange at PCC and the output of other DERs, while (27) restricts the reserve capacity of ESS. Binary variables $y^{\mathrm{zch}},y^{\mathrm{zdic}}$ in (28) represent the operating status of ESS, and (29) tracks the temporal variation of its state of charge. To maintain the operational security, load shedding is permitted in (30), where $\zeta$ is the consumption satisfaction rate.

In the second phase, MGs will negotiate P2P flexibility services in LFM to eliminate power imbalances under rolling estimates. The compact model is presented as (31)-(37).

$$\max_{\hat{y}}\mathrm{c}_{\mathrm{fs}}^{T}\mathbf{B_m}\hat{y}^{\mathrm{pf}}-[\mathrm{c}_{2\mathrm{m}}^{T}(\hat{y}^{\mathrm{pg+}}+\hat{y}^{\mathrm{pg-}})-\mathrm{c}_{2\mathrm{pc}}^{T}(\hat{y}^{\mathrm{Lp+}}-\hat{y}^{\mathrm{Lp-}})-\mathrm{c}_{\mathrm{fp}}^{T}(\hat{y}^{\mathrm{ppc+}}+\hat{y}^{\mathrm{ppc-}})] \tag{31}$$

$$\hat{y}\in\left\{\begin{matrix}\hat{y}^{\mathrm{pf}},\hat{y}^{\mathrm{qf}},\hat{y}^{\mathrm{pg+}},\hat{y}^{\mathrm{pg-}},\hat{y}^{\mathrm{qg}},\hat{y}^{\mathrm{ppc+}},\\ \hat{y}^{\mathrm{ppc-}},\hat{y}^{\mathrm{qpc}},\hat{y}^{\mathrm{s}},\hat{y}^{\mathrm{Lp+}},\hat{y}^{\mathrm{Lp-}},\hat{y}^{\mathrm{Lq}}\end{matrix}\right\} \tag{32}$$

$$([\![y^{\mathrm{ppc*}}]\!]+[\![\hat{x}^{\mathrm{ppc+}}]\!]-[\![\hat{x}^{\mathrm{ppc-}}]\!]+\hat{y}^{\mathrm{ppc+}}-\hat{y}^{\mathrm{ppc-}})+\mathbf{A_m}(y^{\mathrm{pg}}+\hat{y}^{\mathrm{pg+}}-\hat{y}^{\mathrm{pg-}})+\mathbf{B_m}\hat{y}^{\mathrm{pf}}=\check{\mathrm{P}}^{\mathrm{LM}}-([\![y^{\mathrm{Lp*}}]\!]+\hat{y}^{\mathrm{Lp+}}-\hat{y}^{\mathrm{Lp-}}), \tag{33.a}$$

$$(\llbracket y^{\mathrm{qpc}*}\rrbracket+\llbracket\hat{x}^{\mathrm{qpc}}\rrbracket+\hat{y}^{\mathrm{qpc}})+\mathbf{C_m}(y^{\mathrm{qg}}+\hat{y}^{\mathrm{qg}})+\mathbf{B_m}\hat{y}^{\mathrm{qf}}=\breve{\mathrm{Q}}^{\mathrm{LM}}-(\llbracket y^{\mathrm{Lq}*}\rrbracket+\hat{y}^{\mathrm{Lq}}), \tag{33.b}$$

$$-\rho\le\llbracket y^{\mathrm{ppc}*}\rrbracket+\llbracket\hat{x}^{\mathrm{ppc+}}\rrbracket-\llbracket\hat{x}^{\mathrm{ppc-}}\rrbracket+\hat{y}^{\mathrm{ppc+}}-\hat{y}^{\mathrm{ppc-}}+\mathbf{B_m}\hat{y}^{\mathrm{pf}}\le\rho, \tag{34.a}$$

$$-\vartheta\le\llbracket y^{\mathrm{qpc}*}\rrbracket+\llbracket\hat{x}^{\mathrm{qpc}}\rrbracket+\hat{y}^{\mathrm{qpc}}+\mathbf{B_m}\hat{y}^{\mathrm{qf}}\le\vartheta, \tag{34.b}$$

$$0\le\hat{y}^{\mathrm{pg+}}\le\overline{y}^{\mathrm{pg}}, \tag{35.a}$$

$$0\le\hat{y}^{\mathrm{pg-}}\le y^{\mathrm{pg}}, \tag{35.b}$$

$$0\le\mathbf{L_m}(y^{\mathrm{pg}}+\hat{y}^{\mathrm{pg+}}-\hat{y}^{\mathrm{pg-}})\le\mathbf{M_m}y_{it}^{\mathrm{zch}}+\mathbf{N_m}y_{it}^{\mathrm{zdic}}+\mathbf{O_m}\breve{P}_{it}^{\mathrm{gM}}+\varpi, \tag{35.c}$$

$$\mathbf{P_m}\hat{y}^{\mathrm{s}}+\mathbf{Q_m}(y^{\mathrm{pg}}+\hat{y}^{\mathrm{pg+}}-\hat{y}^{\mathrm{pg-}})=\sigma, \tag{36}$$

$$(1-\zeta)\breve{\mathrm{P}}^{\mathrm{LM}}\le\breve{\mathrm{P}}^{\mathrm{LM}}-(\llbracket y^{\mathrm{Lp}*}\rrbracket+\hat{y}^{\mathrm{Lp+}}-\hat{y}^{\mathrm{Lp-}})\le\breve{\mathrm{P}}^{\mathrm{LM}}, \tag{37.a}$$

$$\hat{y}^{\mathrm{Lq}}=(\hat{y}^{\mathrm{Lp+}}-\hat{y}^{\mathrm{Lp-}})\tan\theta. \tag{37.b}$$

The flexibility contracts and redispatch schemes of MGs in (32) are developed by pursuing intra-day profits. The flexibility revenue in (31) is formed by deducting intra-day operating costs from the flexibility service income. (33) restricts the power balance within MGs in rolling windows, and (34) represents the operating boundaries of total power exchange at PCC. (35)-(36) limit the operation condition of the reserve ESS, while (37) is the load shedding restriction when updating intra-day demand $\breve{\mathrm{P}}^{\mathrm{LM}}, \breve{\mathrm{Q}}^{\mathrm{LM}}$. It is noted that P2P trading between MGs is regulated by DSO to some extent through the introduction of flexibility adjustments in (33)-(34), i.e.$\llbracket\hat{x}^{\mathrm{ppc+}}\rrbracket, \llbracket\hat{x}^{\mathrm{ppc-}}\rrbracket, \llbracket\hat{x}^{\mathrm{qpc}}\rrbracket$. The power exchanges at PCC contain the first-stage energy bids submitted by MGs $\llbracket y^{\mathrm{ppc}*}\rrbracket, \llbracket y^{\mathrm{qpc}*}\rrbracket$, the import and export P2P flexibility services $\mathbf{B_m}\hat{y}^{\mathrm{qf}}$ , and the needed flexibility adjustments formulated by MGs and DSO, i.e. $\llbracket\hat{x}^{\mathrm{ppc+}}\rrbracket-\llbracket\hat{x}^{\mathrm{ppc-}}\rrbracket$ and $\hat{y}^{\mathrm{ppc+}}-\hat{y}^{\mathrm{ppc-}}$, respectively, while the flexibility summation transferred to DSO is denoted as $\llbracket y^{\mathrm{pf}}\rrbracket=\hat{y}^{\mathrm{ppc+}}-\hat{y}^{\mathrm{ppc-}}+\mathrm{B_m}\hat{y}^{\mathrm{pf}}$.

## IV. Chance-Constrained Uncertain Distribution Locational Marginal Pricing Formation

The CC-UDLMP is applied in the first stage to quantify the impact of probabilistic uncertainties on secure operations by accounting for the cost of preventive measures. This spatial-temporal differentiated price attracts demand response from MGs to alleviate operating risks caused by uncertainty. As the allocation fee of reserve is not contained, the related variables in (9) are ignored in the formation of CC-UDLMP. Following steps are applied to equivalently reformulate chance constraints

For convenience, the probabilistic system state variable $\tilde{x}^{\mathrm{df}}$, named as $\tilde{\chi}$ for brevity, is denoted as the sum of the original state value $\chi$ obtained under the typical scenario $\bar{\mathrm{P}}_{ut}$ and the offset $\Delta\chi$ caused by uncertain disturbance $\hat{P}_{ut}$. Considering wind and PV DERs are dispatchable within the available range, the indirect influence of $\tilde{P}^{\mathrm{G}}$ on network power flow is reasonably ignored. For simplicity,$\Delta\chi$ is obtained by multiplying the sensitivity matrix $\boldsymbol{S}$ with volatile nodal demands in (38).

$$\tilde{\chi}=\chi+\Delta\chi=\chi+\boldsymbol{S}\cdot\left[\hat{P}^{\mathrm{L}},\hat{Q}^{\mathrm{L}}\right]^T \tag{38}$$

Therefore, (38) is transformed into (39) based on the branch flow model, where $\boldsymbol{\mathcal{J}}$ denotes the inverse matrix $\boldsymbol{S^{-1}}$ of $\boldsymbol{S}$. The right-term state offset column denote the squared nodal voltage, the squared current, and the active and reactive feeder power.

$$\begin{bmatrix}\hat{P}^{\mathrm{L}}\\ \hat{Q}^{\mathrm{L}}\\ 0\\ 0\end{bmatrix}=\boldsymbol{S^{-1}}\Delta x^{\mathrm{df}}=\boldsymbol{\mathcal{J}}\cdot\begin{bmatrix}\Delta x^{v}\\ \Delta x^{l}\\ \Delta x^{p}\\ \Delta x^{q}\end{bmatrix} \tag{39}$$

For brevity, variables in (39) are separately grouped as $\hat{\theta}_U=\left[\hat{P}^{\mathrm{L}},\hat{Q}^{\mathrm{L}}\right]^T$, $\Delta\chi_1=[\Delta x^v,\Delta x^l]^T$ and $\Delta\chi_2=[\Delta x^p,\Delta x^q]^T$. Then this equation can be equivalently converted to (40), where $\mathcal{J}$ denotes the element of $\boldsymbol{\mathcal{J}}$ corresponding to offset column.

$$\begin{bmatrix}\hat{\theta}_U\\ 0\end{bmatrix}=\boldsymbol{\mathcal{J}}\cdot\begin{bmatrix}\Delta\chi_1\\ \Delta\chi_2\end{bmatrix}=\begin{bmatrix}\mathcal{J}_{\mathrm{R}} & \mathcal{J}_{\mathrm{C}}\\ \mathcal{J}_{\mathrm{Cq}} & \mathcal{J}_{\mathrm{v}}\end{bmatrix}\cdot\begin{bmatrix}\Delta\chi_1\\ \Delta\chi_2\end{bmatrix} \tag{40}$$

This matrix calculation is then expanded as the following:

$$\begin{cases}\hat{\theta}_U=\mathcal{J}_{\mathrm{R}}\Delta\chi_1+\mathcal{J}_{\mathrm{C}}\Delta\chi_2\\ 0=\mathcal{J}_{\mathrm{Cq}}\Delta\chi_1+\mathcal{J}_{\mathrm{v}}\Delta\chi_2\end{cases} \tag{41}$$

By substituting the detailed primary column, the offset of the system state variable is obtained as (42), where $\boldsymbol{S_1},\boldsymbol{S_2}$ are related sensitivity matrix of these two column variables.

$$\begin{cases}\Delta\chi_1=\mathcal{J}_{\mathrm{Cq}}^{-1}\mathcal{J}_{\mathrm{v}}(\mathcal{J}_{\mathrm{R}}\mathcal{J}_{\mathrm{Cq}}^{-1}\mathcal{J}_{\mathrm{v}}-\mathcal{J}_{\mathrm{C}})^{-1}\hat{\theta}_U=\boldsymbol{S_1}\hat{\theta}_U\\ \Delta\chi_2=(\mathcal{J}_{\mathrm{C}}-\mathcal{J}_{\mathrm{R}}\mathcal{J}_{\mathrm{Cq}}^{-1}\mathcal{J}_{\mathrm{v}})^{-1}\hat{\theta}_U=\boldsymbol{S_2}\hat{\theta}_U\end{cases} \tag{42}$$

Based on the responsive principle of uncertainty, the chance constraints in (8) and (25) are named as chance state constraint (CSC) and chance decision constraint (CDC) correspondingly. Uncertainties diffusely affect the offset variable by the entire topology in CSC, while the feasible domain in CDC is directly formed by uncertainties. In (43), $S_k$ denotes the coefficient column of the *k*-th state variable $\chi_k$ in sensitivity matrix.

**Chance state constraints:**

$$\begin{cases}\mathrm{Prob}\big(\chi_k+S_k\hat{\theta}_U\ge\underline{\chi}_k\big)\ge 1-\gamma\\ \mathrm{Prob}\big(\chi_k+S_k\hat{\theta}_U\le\overline{\chi}_k\big)\ge 1-\gamma\end{cases}, \tag{43.a}$$

**Chance decision constraints:**

$$\mathrm{Prob}\big(\chi^G\le\bar{P}_u+\hat{P}_u\big)\ge 1-\gamma. \tag{43.b}$$

By substituting the known uncertainty distributions in (1), these chance constraints can be equivalently reformulated into second-order cones-based formulation [23]. In terms of CSC, (43.a) is replaced by its contraposition. For instance, the lower-bound chance constraint is transformed to $\mathrm{Prob}\big(\chi_k+S_k\hat{\theta}_U\le\underline{\chi_k}\big)\le\gamma$. By inserting mean value $\mu_u$ and covariance matrix $\Sigma_u$ in (1) and recasting this inequality into (44), the entire left side of state offset $S_k\hat{\theta}_U$ follows Gaussian distribution with zero mean and unit variance.

$$\mathrm{Prob}\left(\frac{S_k\hat{\theta}_U-S_k\mu_u}{\sqrt{S_k\Sigma_uS_k}}\le\frac{\underline{\chi}-\chi_k-S_k\mu_u}{\sqrt{S_k\Sigma_uS_k}}\right)\le\gamma \tag{44}$$

Hence, this constraint is simplified as $\frac{\underline{\chi}-\chi_k-S_k\mu_u}{\sqrt{S_k\Sigma_uS_k}}\le\Phi^{-1}(\gamma)$. It can then be rewritten to a normal bounding box with revised limits, i.e. $\chi_k\ge\underline{\chi}-S_k\mu_u-\Phi^{-1}(\gamma)\sqrt{S_k\Sigma_uS_k}$. Analogously, the chance constraints are reformulated as (45.a) and (45.b) respectively, where the star superscript denotes the latest limit.

**Chance state constraints:**

$$\begin{cases}\chi_k\ge\underline{\chi}_k^*=\underline{\chi}_k-S_k\mu_u-\Phi^{-1}(\gamma)\left\|S_k\Sigma_u^{\frac{1}{2}}\right\|\\ \chi_k\le\overline{\chi}_k^*=\overline{\chi}_k-S_k\mu_u-\Phi^{-1}(1-\gamma)\left\|S_k\Sigma_u^{\frac{1}{2}}\right\|\end{cases} \tag{45.a}$$

**Chance decision constraints:**

$$\chi^G\le\bar{\mathrm{P}}_u^*=\bar{\mathrm{P}}_u+\mu_u+\Phi^{-1}(\gamma)\left\|S_k\Sigma_u^{\frac{1}{2}}\right\|, \tag{45.b}$$

As can be seen from (45.a), the feasible domain of the system state variable is shrunk to cope with fluctuating demands, which is a precautionary strategy for the system in face of estimated nodal disturbances. Whereas the impact of uncertainty on CDC in (45.b) is solely related to the availability of renewable energy. It is worth noting that while the chance-constrained economic dispatch problems are now transformed to MISOCP models, a large number of integer variables still challenges the formation of CC-UDLMP in subsection III.A. Based on the conclusion in [24], binary variables associated with ESS can be eliminated under cost considerations, but numerous binary directional variables make the feasibility domain yet highly dispersed. Therefore, we performed recursive calculations to obtain the CC-UDLMP in the convex dual counterpart of (46.a). The detailed formulation can be found in Appendix C.

## V. Fully Distributed Iterative Algorithm for Interactive Two-stage Retail Market

### A. Procedure of Fully Distributed Iterative Algorithm

Since the decisions of DSO and MGs are interoperated and the first-stage results directly affect the feasible domain of the second phase, the proposed market mechanism is essentially a two-stage sequential game that reaches equilibrium through limited interactions, as listed in (46)-(47).

**DSO (1st stage):**

$$\begin{cases} \text{primary model: } (x_{k+1}, z_{k+1}) = \mathcal{R}(y_k) \\ \text{dual model: } \tau^{\mathrm{p}}_{k+1} = \mathcal{R}^{+}(y_k, z_{k+1}) \end{cases} \quad (46.a)$$

**Microgrids (1st stage):** $y_{k+1} = \mathcal{R}_b(\tau^{\mathrm{p}}_{k+1})$ (46.b)

**Microgrids (2nd stage):** $\hat{y}_{l+1} = \acute{\mathcal{R}}_b(\hat{x}_l, y)$ (47.a)

**DSO (2nd stage):** $\hat{x}_{l+1} = \acute{\mathcal{R}}(x, y, \hat{y}_{l+1}, z)$ (47.b)

Among this, $x_{k+1}, z_{k+1}, \tau^{\mathrm{p}}_{k+1}$ denote continuous network-side energy dispatch variable set, binary direction variable set and the CC-UDLMP at the (*k+1*)-th round of the first-stage game, respectively, while $y_{k+1}$ corresponds to the energy scheduling scheme of MGs. Analogously, $\hat{x}_{l+1}, \hat{y}_{l+1}$ are regulation output of DSO and MGs at the (*l+1*)-th round of the second-stage game, where day-ahead decisions $x, y, z$ are fixed in solving procedures. $\mathcal{R}(\cdot)$ represents the reaction function of each entity.

With privacy concerns, we propose a fully distributed iterative algorithm that incorporates a consensus-based algorithm to overcome the computational burden imposed by the large-scale integer variables $z$ in primary model of (46.a). The procedure of this algorithm is briefly summarized as follows:

**Algorithm 1: Fully distributed iterative algorithm**

**1st stage-Initiation:** Initialize the iteration label as $k = 0$ and the energy bids of MGs at PCC as $y_k = y_0$.

**Step 1**: Obtain $x_{k+1}, z_{k+1}$ by the primary model of (46.a) with $y_k$ using *the consensus-based distributed algorithm* in the next Subsection, and update $z_{k+1}$ in Step 2. If $k \geq 2$, go to Step 4.

**Step 2:** Obtain $\tau^{\mathrm{p}}_{k+1}$ by the dual model of (46.a) with $y_k, z_{k+1}$, and send it to Step 3.

**Step 3:** Obtain $y_{k+1}$ by (46.b) with $\tau^{\mathrm{p}}_{k+1}$ and go to Step 1.

**Step 4:** *Stopping criteria:* If $|\tau^{\mathrm{p}}_{k+1} - \tau^{\mathrm{p}}_{k}| \leq \varepsilon$ is satisfied, then terminate this first-stage game and go to Step 5; Otherwise, update iteration label $k = k + 1$ and return to Step 1.

**2nd stage-Rolling initiation:** Initialize the rolling step as $s = 1$, and fix the first-stage result $(x, y, z)$ in (47).

**Step 5**: *Initialization:* Update the iteration label as $l = 0$ along with flexibility orders of MGs as $\hat{x}_l = \hat{x}_0$, and go to Step 6.

**Step 6:** Obtain $\hat{y}_{l+1}$ by (47.a) with $\hat{x}_l$ and send it to Step 7.

**Step 7:** Obtain $\hat{x}_{l+1}$ by (47.b) with $\hat{y}_{l+1}$ and send it to Step 8.

**Step 8:** *Stopping criteria:* If $|\hat{y}_{l+1} - \hat{x}_{l+1}| \leq \varepsilon$ is met, then update $s = s + 1$ and turn to Step 9; Otherwise, update $l = l + 1$ and go to Step 6.

**Step 9:** *Rolling stopping criteria:* If $s > 24$, terminate the whole calculation; Otherwise, go to Step 5.

In contrast to the first-stage stopping criterion that requires CC-UDLMP to converge to a fixed point, it is noted that the second stage entails the flexibility adjustments close to the same value, i.e. neither DSO nor MGs may adjust any further. Due to space restrictions, a detailed proof of the existence of the equilibrium solution is provided in Appendix D.

### B. Consensus-based Distributed Algorithm for MISOCP

Given the non-polynomial computability of (46.a), a novel distributed algorithm is used to solve this MISOCP efficiently. Though ADMM is a mature technique to handle large-scale optimizations, its convergence can only be ensured for convex models [25]. Inspired by the Lp-box ADMM algorithm which was designed for the general binary optimization in [26], we improved this method to obtain a stationary solution with acceptable accuracy in MISOCP model.

The primal model of (46.a) is expanded as (48)-(52), where the regular letter such as A represents coefficient matrix and $x, z$ denote continuous and binary variable sets respectively. The variable set $z$ acts as the consensus variable in this model.

$$\min_{x,z} f(x,z) = \mathrm{c}^T x \quad (48)$$

$$\textit{subject to.} \quad \mathrm{A}x = b, \quad (49)$$

$$\mathrm{D}x + \mathrm{E}z \geq f, \quad (50)$$

$$|\mathrm{G}x| \preccurlyeq 0, \quad (51)$$

$$z \in \{0,1\}. \quad (52)$$

The n-dimensional binary variables in (52) are equivalently replaced by the intersections between a box with a (n-1)-dimensional sphere, indicated as $z \in \{0,1\}^n \Longleftrightarrow z \in [0,1]^n \cap \{z: \|z - \frac{1}{2}\mathrm{I}_n\|_p^p = \frac{n}{2^p}\}$, where $p$ is the coefficient of Euclidean norm and $p = 2$ therein. Then the MISOCP can be reformulated into a consensus-based problem as (53)-(58). Notably, $z_1, z_2$ are new consensus variables and $\sigma_1, \sigma_2$ are dual variables.

$$\min_{x,z,z_1,z_2} f(x,z) = \mathrm{c}^T x \quad (53)$$

$$\textit{subject to.} \quad (x,z) \in S_n = \{(x,z): (49) - (51)\}, \quad (54)$$

$$z = z_1 : \sigma_1, \quad (55)$$

$$z = z_2 : \sigma_2, \quad (56)$$

$$z_1 \in S_b = \{z_1 : z_1 \in [0,1]^n\}, \quad (57)$$

$$z_2 \in S_p = \left\{ z_2 : \left\| z_2 - \frac{1}{2}\mathrm{I}_n \right\|_2^2 = \frac{n}{2^2} \right\}. \quad (58)$$

The augmented Lagrangian function of this consensus problem is formed in (59), where the constraints of consensus variables are eliminated for brevity.

$$\begin{aligned} \mathcal{L}(x, z, z_1, z_2, \sigma_1, \sigma_2) = f(x,z) + \sigma_1^T (z - z_1) + \\ \sigma_2^T (z - z_2) + 0.5\rho_1 \|z - z_1\|_2^2 + 0.5\rho_2 \|z - z_2\|_2^2 \end{aligned} \quad (59)$$

The procedure of this improved consensus-based Lp-box ADMM algorithm is summarized as below:

**Algorithm 2: Lp-box ADMM distributed algorithm**

**Initialize** the iteration label as $i = 0$ and parameters as $z_1^i = z_1^0, z_2^i = z_2^0, z^i = z^0, \sigma_1^i = \sigma_1^0, \sigma_2^i = \sigma_2^0, \rho_1^i = \rho_1^0, \rho_2^i = \rho_2^0$.

**Step 1:** Obtain $z_1^{i+1}$ by (60.a) with $z^i, \sigma_1^i, \rho_1^i$ and send it to Step 3 and Step 4. Go to Step 2.

**Step 2:** Obtain $z_2^{i+1}$ by (60.b) with $z^i, \sigma_2^i, \rho_2^i$ and send it to Step 3 and Step 4. Go to Step 3.

**Step 3:** Obtain $z^{i+1}$ by (61) with $z_1^{i+1}, z_2^{i+1}, \sigma_1^i, \sigma_2^i, \rho_1^i, \rho_2^i$ and send it to Step 4. Switch to Step 4.

**Step 4:** Update dual variables $\sigma_1^{i+1}, \sigma_2^{i+1}$ by (62), and obtain the primary residual $R_{po}^{i+1}$ by (63). Switch to Step 5.

**Step 5:** Stopping criteria: If $i \geq i_{pe}$, update $\rho_1^{i+1}, \rho_2^{i+1}$ by (64). If $\max|R_{po}^{i+1}|_\infty \leq \varepsilon$ is satisfied, then terminate the iterative calculation; Otherwise, set $i = i + 1$ and return to Step 1.

The calculation of $z_1^{i+1}$ and $z_2^{i+1}$ are performed in (60), where $\mathrm{Proj}(\bullet)$ stands for the projection of the value in brackets outside the limit between [0,1] to the nearest boundary.

$$z_1^{i+1} = \mathrm{Proj}(z^i + \sigma_1^i/\rho_1^i) \tag{60.a}$$

$$z_2^{i+1} = \frac{n^{1/2}}{2}\frac{z^i+\sigma_1^i/\rho_2^i-\frac{1}{2}\mathrm{I}^n}{\|z^i+\sigma_1^i/\rho_2^i-\frac{1}{2}\mathrm{I}^n\|_2} + \frac{1}{2}\mathrm{I}^n \tag{60.b}$$

By substituting $z_1^{i+1}, z_2^{i+1}, \sigma_1^i, \sigma_2^i, \rho_1^i, \rho_2^i$, the relaxed SOCP-based formulation with penalty terms is listed as (61).

$$\min_{x,z} c^T x + {\sigma_1^i}^T(z - z_1^{i+1}) + {\sigma_2^i}^T(z - z_2^{i+1}) + 0.5\rho_1^i\|z - z_1^{i+1}\|_2^2 + 0.5\rho_2^i\|z - z_2^{i+1}\|_2^2 \tag{61.a}$$

$$subject\ to.\quad (x, z) \in S_n = \{(x, z)\text{: } (49) - (51)\} \tag{61.b}$$

The ascent of multipliers at the ($i$+*1*)-th round is executed by (62). The primary residual is obtained by (63), where the minimal threshold $\omega$ is set as 1e-3.

$$\sigma_1^{i+1} = \sigma_1^i + \rho_1^i(z^{i+1} - z_1^{i+1}) \tag{62.a}$$

$$\sigma_2^{i+1} = \sigma_2^i + \rho_2^i(z^{i+1} - z_2^{i+1}) \tag{62.b}$$

$$R_{po}^{i+1} := \frac{[\|z - z_1^{i+1}\|_2^2, \|z - z_2^{i+1}\|_2^2]}{\max\{\|z - z_1^{i+1}\|_2^2, \omega\ \}} \tag{63}$$

To accelerate the calculation, the penalty parameters in (61) are multiplied by a positive constant number $\alpha$ greater than 1 after the predefined round as below:

$$\rho_1^{i+1} = \alpha\rho_1^i, \rho_2^{i+1} = \alpha\rho_2^i \tag{64}$$

In terms of convergence, since the linear objective function of MISOCP is quadratically differentiable and semi-algebraic with a bounded Hessian matrix, the solution of the subproblem is always exact, and the key steps in this improved application remain unchanged, the convergence of the improved Lp-box algorithm can be ensured based on the proposition in [26].

## VI. Case Studies

A modified IEEE 33-bus system and an IEEE 123-bus system [27] are adopted to demonstrate the proposed market mechanism. As shown in Fig. 4, four MGs are accessed in the 33-bus test feeder, as well as two ESSs, two wind farms (WG) and two PV plants (VG) with installed capacities of 1 MW, 1 MW and 0.5 MW, respectively. It is worth noting that the high share of renewable energy can supply the load demand most of the time, except for the evening peak period. Main parameters of test cases are summarized in Table I. The wind speed and solar irradiance data are obtained from the dataset provided by National Renewable Energy Laboratory [28], and the locational marginal price of the substation is derived from the PJM market [29]. To verify the effectiveness and versatility of the proposed market mechanism, the following cases are designed.

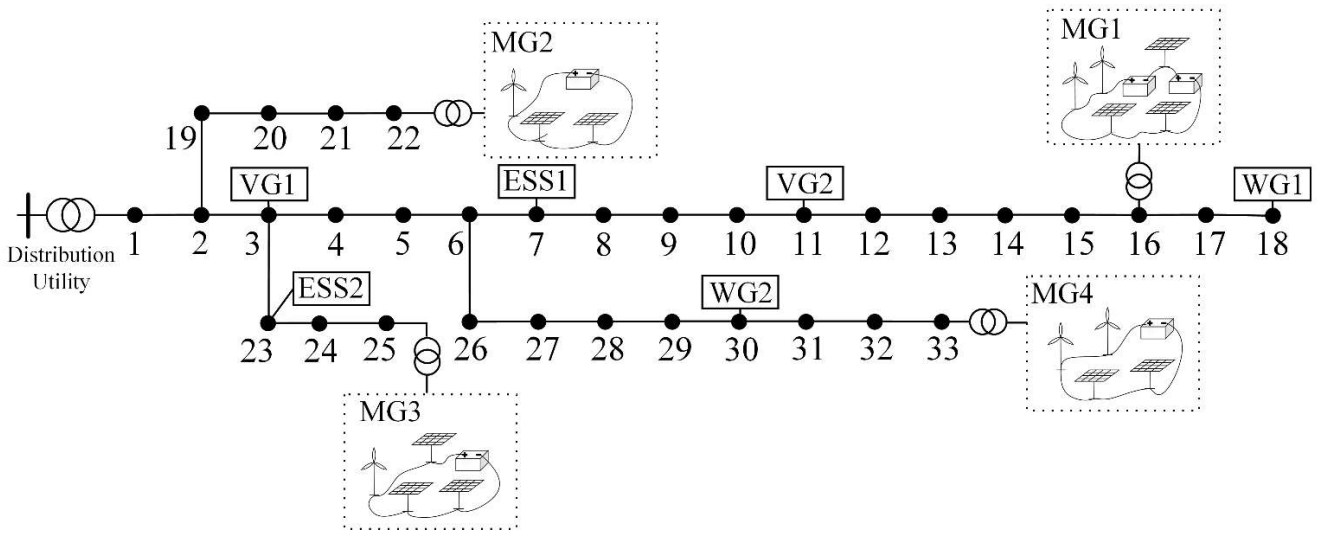

Fig. 4. Modified IEEE 33-bus distribution system

TABLE I

Main parameters in case study

| Parameter | Value | Parameter | Value | Parameter | Value |
|---|---|---|---|---|---|
| $c^{nl}$ | 15\$/MWh | $c_{2d}$ | 30\$/MWh | $\zeta$ | 1 |
| $c_{1pc}$ | 30\$/MWh | $c_{2m}$ | 10\$/MWh | $\alpha$ | 1.2 |
| $c_{2pc}$ | 50\$/MWh | $c_{fp}$ | 50\$/MWh | $\tan\theta$ | 0.95 |
| $c_{ppc}$ | 50\$/MWh | $c_{fs}$ | 20\$/MWh | $i_{pe}$ | 5 |

TABLE II

Cumulative results of the first-stage day-ahead energy trading

| Entity | Operating cost/\$ | Power injection/MW | Wind energy/MW | PV energy/MW |
|---|---|---|---|---|
| DSO | 586.15 | 5.05 | 26.00 | 8.53 |
| MG1 | -52.19 | / | 3.45 | 2.62 |
| MG2 | -4.91 | / | 1.37 | 1.66 |
| MG3 | -14.16 | / | 2.08 | 2.62 |
| MG4 | -36.32 | / | 2.74 | 1.66 |
| **Entity** | **Charging energy/MW** | **Discharging energy/MW** | **Load shedding/MW** | **Energy exchange/MW** |
| DSO | 1.26 | 1.14 | / | / |
| MG1 | 0.17 | 0.16 | 0.41 | -5.38 |
| MG2 | 0.11 | 0.10 | 0.61 | -2.02 |
| MG3 | 0.06 | 0.06 | 1.43 | -2.34 |
| MG4 | 0.06 | 0.06 | 0.41 | -3.73 |

(1) Benchmark case: Discussed in Subsections VI.A, B, and C to provide a benchmark for comparative analysis. It gives a comprehensive view of the proposed market mechanism.

(2) DLMP case: Discussed in Subsection VI.B to illustrate the effectiveness of the proposed CC-UDLMP in pre-empting uncertainties. This case adopts the typical scenario to model the first-stage uncertainty and correspondingly forms deterministic DLMP to clear energy transactions.

(3) No P2P case: Discussed in Subsection VI.C to examine the role of P2P flexibility sharing in improving market efficiency. This case replaces the second-stage P2P sharing pattern in benchmark case with centralized network-side flexibility service scheduling of DSO.

(4) Large-scale case: Discussed in Subsection VI.D to validate the computational performance and calculational accuracy of the proposed market mechanism. This case is tested on the IEEE 123-bus system based on the proposed mechanism.

### A. Benchmark case

For convenience, the transaction proposals for each phase are discussed in sequence. The day-ahead energy trading results are summarized in Table II. In general, it is observed that renewable energy is heavily employed to supply distribution networks and MGs over the time horizon compared to upstream injection, while the negative cumulative amount of exchanged power also

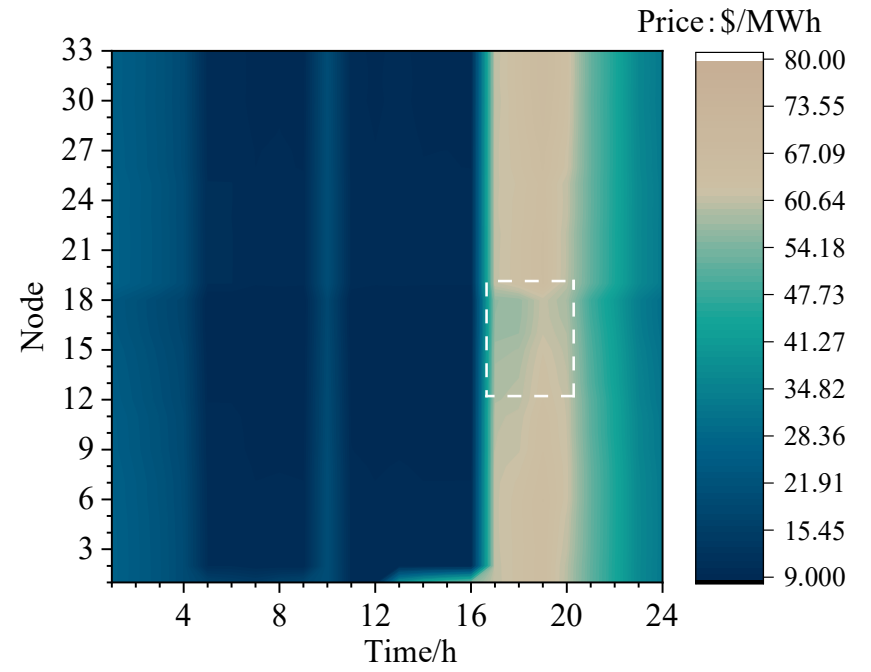

Fig. 5. CC-UDLMP obtained at the first-stage transactions

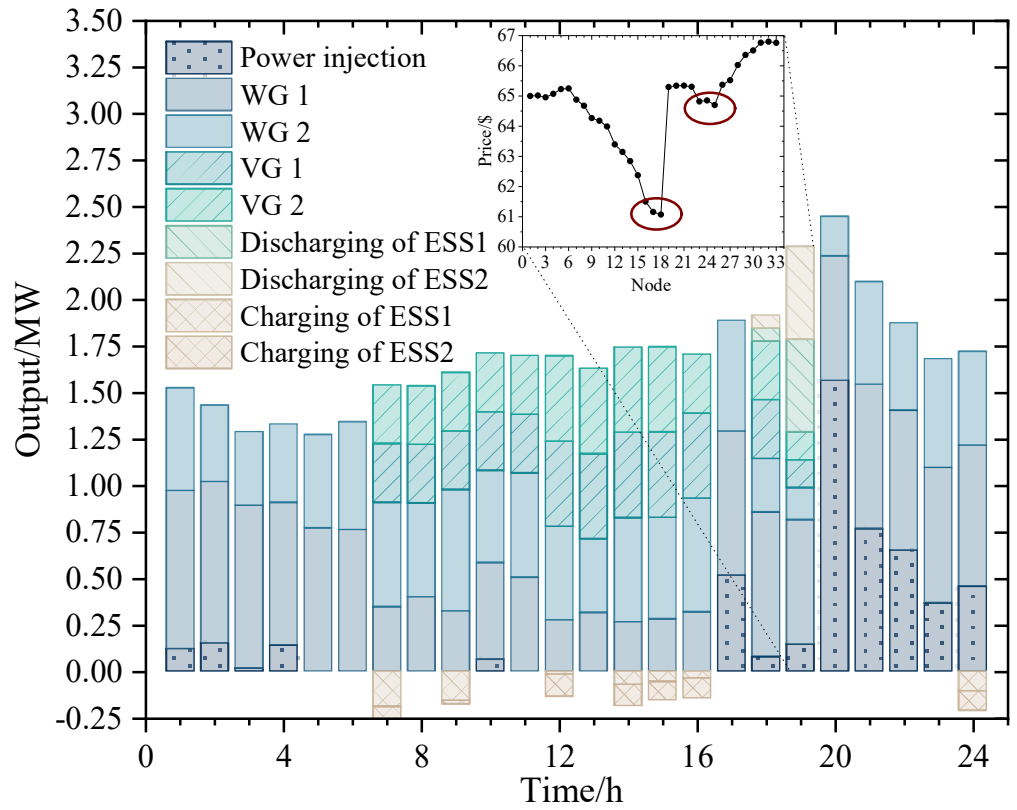

Fig. 6. Day-ahead dispatch scheme of distribution system

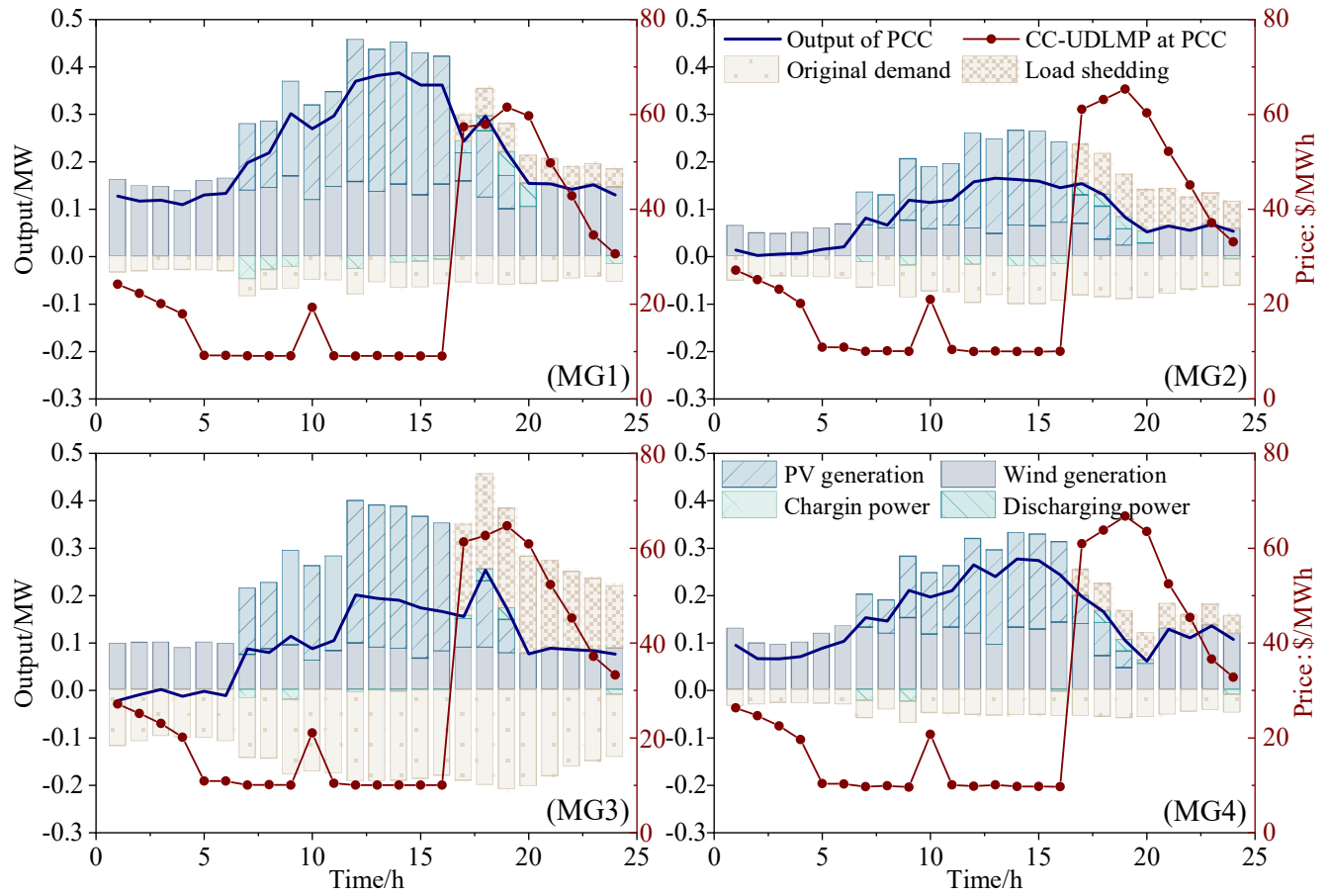

Fig. 7. Day-ahead dispatch schemes and energy exchange profile of MGs

indicates that MGs sold more electricity to DSO than bought. To further analyze the detailed output of all devices at day-ahead clearing prices, the detailed CC-UDLMP and the day-ahead dispatch scheme of DSO are depicted in Fig.5 and Fig.6, respectively. As can be seen from Fig.5 and Fig.6, the CC-UDLMP remains low during the daytime with abundant clean energy, while during night hours from 17:00 to 20:00, the increase in load demand requires power injections from the upstream substation, thus driving up the overall price level. However, it is observed from the enlarged view of CC-UDLMP in Fig.6, the cheaper WGs operating at the available boundary serves as marginal units at the peak hour 19:00, leading to local price pockets in the white box and red circles of Fig.5 and Fig.6, respectively. The day-ahead dispatch schemes and cumulative power exchange profile of MGs are depicted in Fig.7. It shows that, MG operators tend to charge during daytime with lower CC-UDLMP, while discharging and exerting load shedding

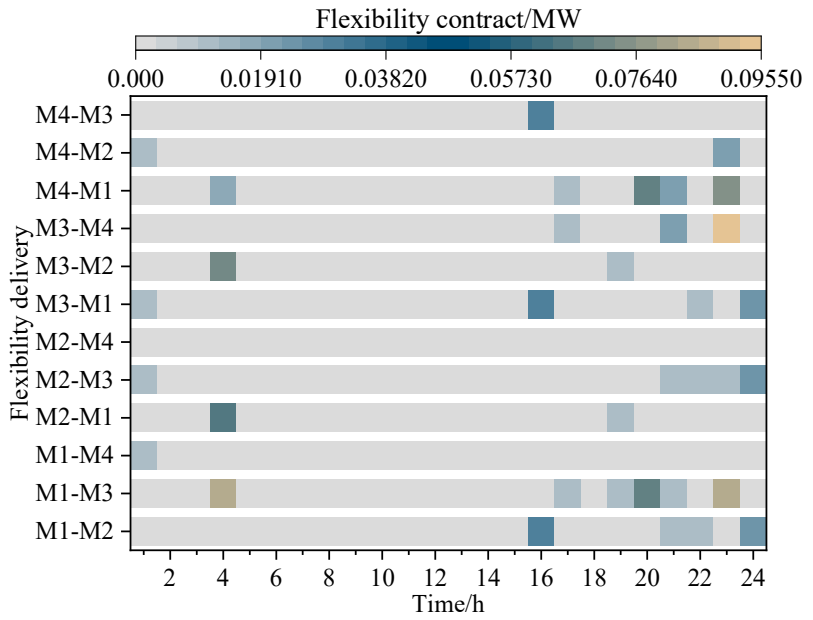

Fig. 8. P2P flexibility contracts between MGs in benchmark case

TABLE III
CUMULATIVE RESULTS OF SECOND-STAGE INTRA-DAY FLEXIBILITY TRADING

| Entity | Operating cost/$ | Power injection/MW | Wind energy/MW | PV energy /MW |
|---|---|---|---|---|
| DSO | 311.33 | +3.40, -0.00 | +2.95, -0.00 | +0.59, -0.00 |
| MG1 | 40.00 | / | +0.00, -0.40 | +0.00, -0.26 |
| MG2 | 21.00 | / | +0.00, -0.17 | +0.00, -0.16 |
| MG3 | 30.90 | / | +0.00, -0.23 | +0.00, -0.26 |
| MG4 | 34.40 | / | +0.00, -0.34 | +0.00, -0.16 |
| **Entity** | **Load shedding/MW** | **Import from DSO/MW** | **Import from P2P trading /MW** | **Export from P2P trading /MW** |
| MG1 | +0.10, -0.03 | +0.60 -0.00 | 0.23 | 0.25 |
| MG2 | +0.16, -0.06 | +0.19, -0.04 | 0.21 | 0.16 |
| MG3 | +0.15, -0.37 | +0.66, -0.04 | 0.30 | 0.20 |
| MG4 | +0.19, -0.03 | +0.47, -0.01 | 0.19 | 0.32 |

during high clearing prices periods for arbitrages.

The intra-day flexibility trading results are summarized in Table III, where symbols +/− denote upward and downward flexibility adjustments respectively. As the power imbalance is relatively trivial compared to energy transactions and ESS has been fully utilized, the reserve capacity and the unchanged operation of ESS are ignored for brevity. It is observed from Table III that all entities spend operating costs to eliminate the imbalanced power caused by rolling estimates. To adapt to the supply shortage of internal DERs, the outputs of wind and PV generations within MGs are decreased and most of daytime load shedding is then increased. In regards to P2P-based flexibility sharing, the reduced usability of internal DERs leads to insignificant net flexibility exchange of individual MG, while as a whole, flexibility eventually shifts from MG 4 to MG 3. In terms of DSO, due to the increased nodal demand in estimates, the flexibility output of networked devices is increased to supply the imbalanced power and generation scarcity of MGs.

To further explore the P2P flexibility transactions between MGs under the regulation of DSO, detailed flexibility contracts are shown in Fig.8. It is observed that the flexibility transactions take place during early morning and night hours, during which the redundant flexibility capacity of WGs in these MGs are economically shared to MGs with flexibility needs. The P2P trading not only improves the overall economic efficiency but also leverages the flexibility potential of end prosumers and relieves the pressure of DSO to uniformly schedule network-side flexibility resources. The detailed impact of P2P flexibility sharing will be further examined in Subsection VI.C.

### B. *Impacts of CC-UDLMP settlement on handling uncertainty*

To investigate the impact of uncertainty pricing on market

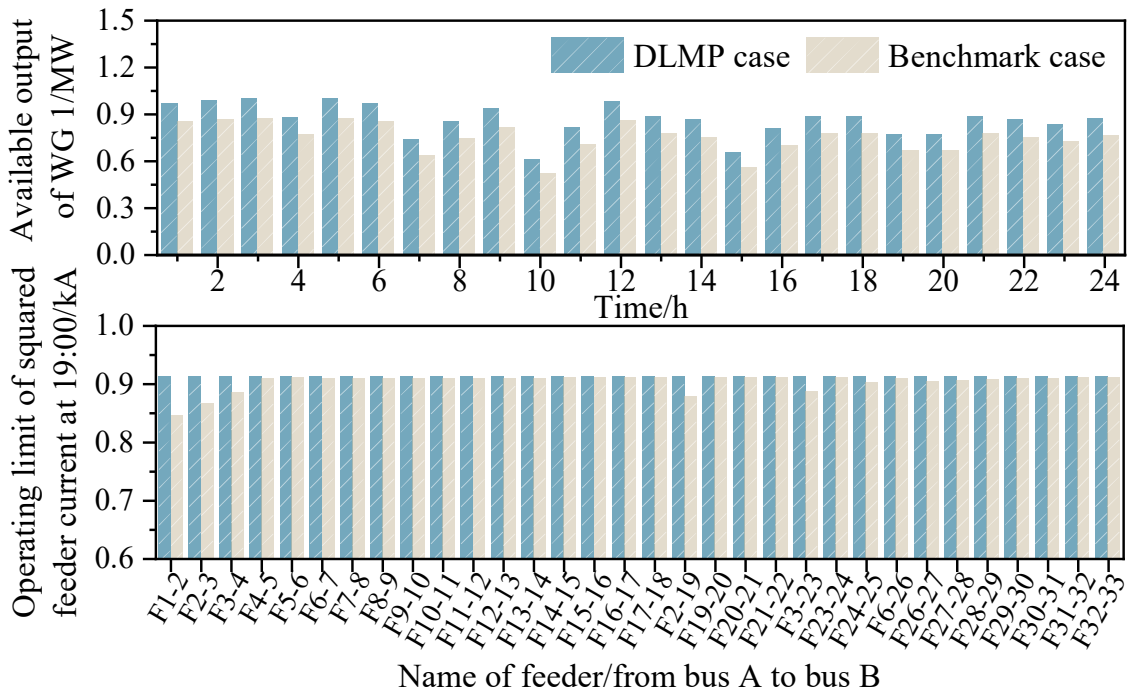

Fig. 9. Operating limits of WG 1 and squared current in two cases

TABLE IV

CUMULATIVE DAY-AHEAD RESULTS BETWEEN BASE CASE AND DLMP CASE

| Cases | Operating cost/$ | Power injection /MW | Wind energy /MW | PV energy /MW |
|---|---|---|---|---|
| DLMP case | 394.93 | 0.82 | 35.49 | 21.56 |
| Benchmark case | 586.15 | 5.05 | 35.63 | 17.10 |

settlement, we first compared the first-stage dispatch schemes and clearing price curves derived from benchmark and DLMP cases, i.e. CC-UDLMP and DLMP, and then further analyzed the final market settlements formed by these two approaches.

For brevity, some of the cumulative energy trading schemes obtained from these cases are summarized in Table IV. With higher availability of WG and VG than benchmark case, the output of VGs in DLMP case is significantly increased and the upstream power is then decreased, leading to savings in operating costs of procuring power injection. To further analyze the updated systematic operational domain in benchmark case, we compared the available output of WG1 and the operating limit of squared feeder current at 19:00 of these two cases in Fig.9. Affected by the precautionary provisions in (45), we noted that the available output of WG1 and the operating limits of squared feeder current are curtailed in benchmark case, while the latter exposes the topological correlation that the shrinkage of feeder operating limits adjacent to power sources is greater than the others. To study the discrepancy between clearing prices in these cases, we plotted topological-based prices at 19:00 in Fig.10, where the circle of various sizes represents the value obtained by subtracting DLMP from CC-UDLMP. It is observed from Fig.10, the closer to the distribution substation and other power supplies, the larger the CC-UDLMP compared to the DLMP. The reason for these observations is that the main feeder needs to retain adequate operating margins to transmit power in contingencies, while the increment of CC-UDLMP along the feeder is due to the increase of network losses.

Additionally, to further examine the differences of the second-stage flexibility adjustments in DLMP and benchmark cases, we summarized the cumulative intra-day flexibility schemes in Table V. It is observed that due to the substantial decline in the availability of clean energy of rolling estimates, the outputs of WG and VG on both network and MGs sides in DLMP case are considerably reduced compared to benchmark case. In this situation, bi-directional flexibility adjustments are intensively applied to maintain power balances in DLMP case, thereby resulting in the increment of intra-day operating costs of both DSO and MGs.

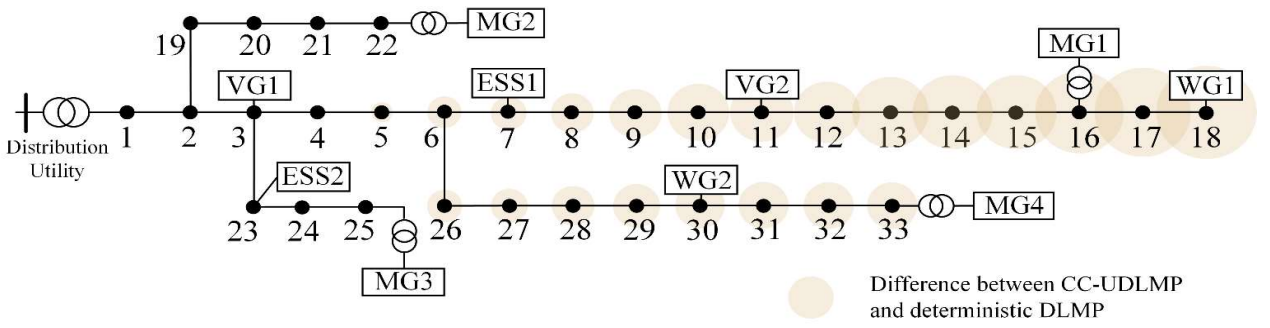

Fig. 10. Price differences between CC-UDLMP and DLMP at 19:00

TABLE V

CUMULATIVE INTRA-DAY FLEXIBILITY SCHEMES IN DLMP AND BASE CASES

| Case | Entity | Power injection[a] /MW | Clean energy[a] /MW | Load shedding[a] /MW | DSO import[a] /MW | 2nd stage operating cost/$ |
|---|---|---|---|---|---|---|
| DLMP case | DSO | +2.58, -0.00 | +1.31, -4.16 | / | / | 458.98 |
| | MGs | / | +0.32, -5.10 | +2.70, -0.15 | +3.67, -1.43 | 295.14 |
| Bench-mark case | DSO | +3.40, -0.00 | +3.54, -0.00 | / | / | 311.33 |
| | MGs | / | +0.00, -1.99 | +0.61, -0.48 | +1.92, -0.09 | 118.76 |

a. The quantity denotes the second-stage bi-directional flexibility adjustment, while flexibility in the table header is ignored for brevity.

TABLE VI

CUMULATIVE INTRA-DAY FLEXIBILITY SCHEMES IN NO P2P CASE

| Case | Entity | Power injection /MW | Clean energy /MW | Load shedding /MW | DSO import /MW | 2nd stage operating cost/$ |
|---|---|---|---|---|---|---|
| No P2P case | DSO | +2.90, -0.00 | +2.85, -0.00 | / | / | 229.56 |
| | MGs | / | +0.00, -1.99 | +1.46, -0.28 | +0.78, -0.00 | 132.69 |

### *C. Role of P2P pattern in flexibility transactions*

To study the role of P2P flexibility trading in improving market efficiency, we concluded the intra-day flexibility dispatch plan derived by no P2P case in Table VI. It is observed that without P2P flexibility trading, MG operators will execute load shedding rather than purchasing upward flexibility from DSO to cope with internal supply shortages, which requires higher operating costs than benchmark case. As for benchmark case, the regulation of DSO that ensures the delivery of P2P flexibility sharing may result in higher network losses, leading to the growth in the operating cost of DSO in practical conditions. However, with redundant flexibility capacity in MGs, the P2P flexibility sharing not only theoretically exploits the flexibility potential of prosumers but also provides an economic way to reduce the allocation of network-side reserve capacity while ensuring system safety in the face of uncertainty.

### *D. Computational performance analysis in large-scale case*

The consensus-based Lp-box ADMM algorithm solves the computational challenges of the probabilistic undirected power flow model, thus improving the generality of the proposed two-stage sequential game in large-scale cases. To examine the applicability of this algorithm in the IEEE 123-bus test system, we compared the computational performance of applying the branch and bound algorithm (BBA) and the improved Lp-box ADMM method in the first-stage probabilistic economic dispatch model of DSO in Table VII. It is observed that the Lp-box ADMM algorithm outperforms BBA in both the computing time and total costs. Due to the highly discontinuous feasible domain of MISOCP, BBA tends to degenerate into exhaustive search, while the improved Lp-box ADMM avoids frequent

TABLE VII

COMPUTING PERFORMANCE IN TRADITIONAL BRANCH AND BOUND METHOD AND IMPROVED LP-BOX ADMM METHOD

| Cases | Computing time/s | Total costs/$ | Maximum error of dispatch scheme | Maximum error of binary variables | Iteration times |
|---|---|---|---|---|---|
| BBA | 2248 | 309.18 | 0 | 0 | 1 |
| Lp-box | 479 | 306.56 | 7% | 8e-5 | 7 |
| Fixed path | 2.60 | 309.71 | 2% | 0 | 1 |

heuristic search for optimal cuts on rooted trees, reflecting the robustness of the solution. However, as the Lp-box ADMM is to find a stationary point rather than the optimal exact solution, an economic dispatch SOCP model with fixed power flow paths derived from the improved Lp-box ADMM algorithm is applied to derive the optimal solution in MISOCP.

## VII. CONCLUSION

An interactive two-stage joint retail market mechanism for multiple MGs is presented to guide the bids of prosumers to adapt to uncertainties. It combines day-ahead energy trading in a centralized market and intra-day flexibility transactions in local flexibility markets between MGs under the supervision of DSO. The time-varying CC-UDLMP price, which considers the cost of precautions against uncertainty, is applied to incentivize MGs to develop profitable energy exchanges while complying with operational instructions of the distribution system. Besides, to exploit the potential of end prosumers, a novel P2P flexibility trading is exerted in LFMs under the supervision of DSO, inserting essential secure operation considerations in flexibility sharing. Moreover, a fully distributed iterative algorithm is proposed to resolve this market mechanism with privacy issues considered, where an improved Lp-box ADMM algorithm is leveraged to address the computational burden caused by probabilistic MISOCP in a distributed manner. It is validated that the proposed market mechanism design not only provides a certain degree of autonomous security but also effectively improves overall market efficiency, which can be further expanded for the market of integrated energy systems.

## APPENDIX A

### MATHEMATIC FORMULATION OF TWO-STAGE MARKET SETTLEMENT MODEL OF DSO

The first-stage chance-constrained energy clearing model is constructed as (A.1)-(A.16).

$$\min_{x,z} \Big\{ \mathrm{c}_{it}^{\mathrm{LMP}} x_{it}^{\mathrm{pg}} + \textstyle\sum_{i\in\Psi_{\mathrm{R}}} \mathrm{c}_i^{\mathrm{pr}} x_{it}^{\mathrm{pr}} + \sum_{i\in\Psi_{\mathrm{B}}} \big(\mathrm{c}_i^{\mathrm{ch}} x_{it}^{\mathrm{ch}} + \mathrm{c}_i^{\mathrm{dic}} x_{it}^{\mathrm{dic}}\big) + \mathrm{c}^{\mathrm{nl}} \sum_{ij\in\Psi_{\mathrm{f}}} \big[r_{ij}\big(\tilde{x}_{ijt}^{\mathrm{l}+} - \tilde{x}_{ijt}^{\mathrm{l}-}\big)\big] \Big\} \tag{A.1}$$

*subject to.*

$$(x,z) \in \left\{ \begin{matrix} x_{it}^{\mathrm{pg}}, x_{it}^{\mathrm{qg}}, x_{it}^{\mathrm{pr}}, x_{it}^{\mathrm{qr}}, x_{it}^{\mathrm{ch}}, x_{it}^{\mathrm{dic}}, x_{it}^{\mathrm{s}}, \tilde{x}_{ijt}^{\mathrm{p}+}, \\ \tilde{x}_{ijt}^{\mathrm{p}-}, \tilde{x}_{ijt}^{\mathrm{q}+}, \tilde{x}_{ijt}^{\mathrm{q}-}, \tilde{x}_{ijt}^{\mathrm{l}+}, \tilde{x}_{ijt}^{\mathrm{l}-}, \tilde{x}_{it}^{\mathrm{v}}, z_{ijt}^{+}, z_{ijt}^{-}, \\ z_{it}^{\mathrm{ch}}, z_{it}^{\mathrm{dic}}, \underline{x}_{it}^{\mathrm{pg}}, \overline{x}_{it}^{\mathrm{pg}}, \underline{x}_{it}^{\mathrm{ch}}, \overline{x}_{it}^{\mathrm{ch}}, \underline{x}_{it}^{\mathrm{dic}}, \overline{x}_{it}^{\mathrm{dic}} \end{matrix} \right\} \tag{A.2}$$

$$x_{it}^{\mathrm{pg}} + x_{it}^{\mathrm{pr}} + x_{it}^{\mathrm{dic}} - x_{it}^{\mathrm{ch}} + \textstyle\sum_{j\in\Omega_i^+}\big(x_{ijt}^{\mathrm{p}} - r_{ij}\tilde{x}_{ijt}^{\mathrm{l}+}\big) + \sum_{j\in\Omega_i^-}\big(r_{ij}\tilde{x}_{ijt}^{\mathrm{l}-} - x_{ijt}^{\mathrm{p}}\big) - [\![y_{it}^{\mathrm{ppc}}]\!] = \tilde{P}_{it}^{\mathrm{L}} : \tau_{it}^{\mathrm{p}}, \tag{A.3a}$$

$$x_{it}^{\mathrm{qg}} + x_{it}^{\mathrm{qr}} + \textstyle\sum_{j\in\Omega_i^+}\big(x_{ijt}^{q} - x_{ij}\tilde{x}_{ijt}^{\mathrm{l}+}\big) + \sum_{j\in\Omega_i^-}\big(x_{ij}\tilde{x}_{ijt}^{\mathrm{l}-} - x_{ijt}^{\mathrm{q}}\big) - [\![y_{it}^{\mathrm{qpc}}]\!] = \tilde{Q}_{it}^{\mathrm{L}} : \tau_{it}^{\mathrm{q}}, \forall i \in \Psi_{\mathrm{N}}, t \in \mathrm{T}. \tag{A.3b}$$

$$\tilde{x}_{it}^{\mathrm{v}} - \tilde{x}_{jt}^{\mathrm{v}} = 2\big(r_{ij}x_{ijt}^{\mathrm{p}} + x_{ij}x_{ijt}^{\mathrm{q}}\big) - \big(r_{ij}^2 + x_{ij}^2\big)\tilde{x}_{ijt}^{\mathrm{l}+} : \omega_{ijt}, \tag{A.4}$$

$$\big|\tilde{x}_{ijt}^{\mathrm{l}+} + \tilde{x}_{it}^{\mathrm{v}}; 2\,x_{ijt}^{\mathrm{p}+}; 2\,x_{ijt}^{\mathrm{q}+}; \tilde{x}_{ijt}^{\mathrm{l}+} - \tilde{x}_{it}^{\mathrm{v}}\big| \preceq 0 : \big|d_{ijt}^{+}; d_{ijt}^{\mathrm{A}+}; d_{ijt}^{\mathrm{B}+}; d_{ijt}^{\mathrm{C}+}\big|, \tag{A.5a}$$

$$\big|-\tilde{x}_{ijt}^{\mathrm{l}-} + \tilde{x}_{jt}^{\mathrm{v}}; -2x_{ijt}^{\mathrm{p}-}; -2x_{ijt}^{\mathrm{q}-}; -\tilde{x}_{ijt}^{\mathrm{l}-} - \tilde{x}_{jt}^{\mathrm{v}}\big| \preceq 0 : \big|d_{ijt}^{-}; d_{ijt}^{\mathrm{A}-}; d_{ijt}^{\mathrm{B}-}; d_{ijt}^{\mathrm{C}-}\big|, \tag{A.5b}$$

$$\mathrm{Prob}\big(0 \le \tilde{x}_{ijt}^{\mathrm{l}+} \le z_{ijt}^{+}\bar{\mathrm{L}}_{ij}\big) \ge 1-\gamma : \lambda_{ijt}^{\mathrm{f}+}, \varphi_{ijt}^{\mathrm{f}+}, \tag{A.6a}$$

$$\mathrm{Prob}\big(0 \le \tilde{x}_{ijt}^{\mathrm{p}+} \le z_{ijt}^{+}\bar{\mathrm{P}}_{ij}\big) \ge 1-\gamma : \lambda_{ijt}^{\mathrm{p}+}, \varphi_{ijt}^{\mathrm{p}+}, \tag{A.6b}$$

$$\mathrm{Prob}\big(-z_{ijt}^{+}\bar{\mathrm{Q}}_{ij} \le \tilde{x}_{ijt}^{\mathrm{q}+} \le z_{ijt}^{+}\bar{\mathrm{Q}}_{ij}\big) \ge 1-\gamma : \lambda_{ijt}^{\mathrm{q}+}, \varphi_{ijt}^{\mathrm{q}+}, \tag{A.6c}$$

$$\mathrm{Prob}\big(-z_{ijt}^{-}\bar{\mathrm{L}}_{ij} \le \tilde{x}_{ijt}^{\mathrm{l}-} \le 0\big) \ge 1-\gamma : \lambda_{ijt}^{\mathrm{f}\text{-}}, \varphi_{ijt}^{\mathrm{f}\text{-}}, \tag{A.7a}$$

$$\mathrm{Prob}\big(-z_{ijt}^{-}\bar{\mathrm{P}}_{ij} \le \tilde{x}_{ijt}^{\mathrm{p}-} \le 0\big) \ge 1-\gamma : \lambda_{ijt}^{\mathrm{p}-}, \varphi_{ijt}^{\mathrm{p}-}, \tag{A.7b}$$

$$\mathrm{Prob}\big(-z_{ijt}^{-}\bar{\mathrm{Q}}_{ij} \le \tilde{x}_{ijt}^{\mathrm{q}-} \le z_{ijt}^{-}\bar{\mathrm{Q}}_{ij}\big) \ge 1-\gamma : \lambda_{ijt}^{\mathrm{q}-}, \varphi_{ijt}^{\mathrm{q}-}, \tag{A.7c}$$

$$\mathrm{Prob}\big(\underline{\mathrm{x}}_{it}^{\mathrm{v}} \le \tilde{x}_{it}^{\mathrm{v}} \le \overline{\mathrm{x}}_{it}^{\mathrm{v}}\big) \ge 1-\gamma : \lambda_{it}^{\mathrm{v}}, \varphi_{it}^{\mathrm{v}}, \forall i \in \Psi_{\mathrm{N}\backslash\{1\}}, \tag{A.8a}$$

$$\tilde{x}_{1t}^{\mathrm{v}} = 1 : \omega_t^{\mathrm{bn}}, \forall t \in \mathrm{T}. \tag{A.8b}$$

$$z_{ijt}^{+} + z_{ijt}^{-} = 1, z_{ijt}^{+}, z_{ijt}^{-} \in \{0,1\}, \forall ij \in \Psi_{\mathrm{f}}, t \in \mathrm{T}. \tag{A.9}$$

$$\mathrm{Prob}\big(0 \le x_{it}^{\mathrm{pr}} \le \tilde{P}_{it}^{\mathrm{g}}\big) \ge 1-\gamma : \lambda_{it}^{\mathrm{pr}}, \varphi_{it}^{\mathrm{pr}}, \tag{A.10}$$

$$0 \le x_{it}^{\mathrm{qr}} \le \overline{\mathrm{Q}}_i^{\mathrm{g}} : \lambda_{it}^{\mathrm{qr}} \ge 0, \varphi_{it}^{\mathrm{qr}} \le 0, \forall i \in \Psi_{\mathrm{R}}, t \in \mathrm{T}. \tag{A.11}$$

$$0 \le x_{it}^{\mathrm{pg}} - \underline{x}_{it}^{\mathrm{pg}} : \lambda_{it}^{\mathrm{pg}}, \tag{A.12a}$$

$$x_{it}^{\mathrm{pg}} + \overline{x}_{it}^{\mathrm{pg}} \le \overline{\mathrm{P}}_{it}^{\mathrm{g}} : \varphi_{it}^{\mathrm{pg}}, \tag{A.12b}$$

$$0 \le x_{it}^{\mathrm{qg}} \le \overline{\mathrm{Q}}_i^{\mathrm{gr}} : \lambda_{it}^{\mathrm{qg}}, \varphi_{it}^{\mathrm{qg}}, \forall i \in \Psi_{\mathrm{G}}, t \in \mathrm{T}. \tag{A.12c}$$

$$0 \le x_{it}^{\mathrm{ch}} - \underline{x}_{it}^{\mathrm{ch}} : \lambda_{it}^{\mathrm{ch}}, \tag{A.13a}$$

$$x_{it}^{\mathrm{ch}} + \overline{x}_{it}^{\mathrm{ch}} \le z_{it}^{\mathrm{ch}}\overline{\mathrm{P}}_i^{\mathrm{ch}} : \varphi_{it}^{\mathrm{ch}}, \tag{A.13b}$$

$$0 \le x_{it}^{\mathrm{dic}} - \underline{x}_{it}^{\mathrm{dic}} : \lambda_{it}^{\mathrm{dic}}, \tag{A.14a}$$

$$x_{it}^{\mathrm{dic}} + \overline{x}_{it}^{\mathrm{dic}} \le z_{it}^{\mathrm{dic}}\overline{\mathrm{P}}_i^{\mathrm{dic}} : \varphi_{it}^{\mathrm{dic}}, \tag{A.14b}$$

$$z_{it}^{\mathrm{ch}} + z_{it}^{\mathrm{dic}} \le 1, z_{it}^{\mathrm{ch}}, z_{it}^{\mathrm{dic}} \in \{0,1\}, \tag{A.15}$$

$$x_{i,t+1}^{\mathrm{s}} = x_{it}^{\mathrm{s}} + \big(x_{it}^{\mathrm{ch}}\eta_{\mathrm{ch}} - x_{it}^{\mathrm{dic}}/\eta_{\mathrm{dic}}\big)\,\Delta t/\mathrm{E}_i : \omega_{it}^{\mathrm{s}}, \tag{A.16a}$$

$$\underline{\mathrm{s}} \le x_{it}^{\mathrm{s}} \le \overline{\mathrm{s}} : \lambda_{it}^{\mathrm{s}}, \varphi_{it}^{\mathrm{s}}, \tag{A.16b}$$

$$x_{i1}^{\mathrm{s}} = s^{\mathrm{ini}} : \omega_i^{\mathrm{ini}}, x_{i,\mathrm{T}+1}^{\mathrm{s}} = s^{\mathrm{end}} : \omega_i^{\mathrm{end}}, \forall i \in \Psi_{\mathrm{B}}, t \in \mathrm{T}. \tag{A.16c}$$

Where $x_{it}^{\mathrm{pg}}, x_{it}^{\mathrm{qg}}, x_{it}^{\mathrm{pr}}, x_{it}^{\mathrm{qr}}$ are the active and reactive outputs of the upstream injection and DERs. $x_{it}^{\mathrm{ch}}, x_{it}^{\mathrm{dic}}$ are charging and discharging power of ESS, while $x_{it}^{\mathrm{s}}$ denotes its state of charge. $\tilde{x}_{ijt}^{\mathrm{p}+}, \tilde{x}_{ijt}^{\mathrm{q}+}, \tilde{x}_{ijt}^{\mathrm{l}+}$ represent probabilistic active power, reactive power and squared current through feeder in the forward power flow solution, while $\tilde{x}_{ijt}^{\mathrm{p}-}, \tilde{x}_{ijt}^{\mathrm{q}-}, \tilde{x}_{ijt}^{\mathrm{l}-}$ correspond to that in the reverse solution, and $\tilde{x}_{it}^{\mathrm{v}}$ denotes probabilistic squared nodal voltage. $z_{ijt}^{+}, z_{ijt}^{-}$ are binary directional variables, and $z_{it}^{\mathrm{ch}}, z_{it}^{\mathrm{dic}}$ are binary state variables of ESS. $\underline{x}_{it}^{\mathrm{pg}}, \overline{x}_{it}^{\mathrm{pg}}, \underline{x}_{it}^{\mathrm{ch}}, \overline{x}_{it}^{\mathrm{ch}}, \underline{x}_{it}^{\mathrm{dic}}, \overline{x}_{it}^{\mathrm{dic}}$ are bi-directional reserve capacity of power injection and ESS. Among them, (A.3)-(A.9) are undirected second-order cone-based power flow model. (A.10)-(A.11) limits the output of DERs, and (A.12) restricts the operation of the upstream injection and its reserve allocation. (A.13)-(A.14) are operational limits of charging and discharging output along with reserve capacity of ESS, while (A.15)-(A.16) restrict the state of charge of ESS. It is worth noting that the system variables $\tilde{x}_{ijt}^{\mathrm{p}}, \tilde{x}_{ijt}^{\mathrm{q}}, \tilde{x}_{ijt}^{\mathrm{l}}, \tilde{x}_{it}^{\mathrm{v}}$ and the active output of DERs are restricted by chance constraints, and the interchange power at PCC is updated by the latest feedback of MG operators, i.e. $[\![y_{it}^{\mathrm{ppc}}]\!]$ and $[\![y_{it}^{\mathrm{qpc}}]\!]$ in (A.3). The variable behind each constraint denotes its Lagrangian multiplier.

The second-stage flexibility regulation settlement model is detailed as (A.17)-(A.29).

$$\begin{aligned}\min_{\hat{x}}\ & \textstyle\sum_{t\in \mathrm{T}_s}\big\{\mathrm{c}_t^{\mathrm{r}}\big[\sum_{i\in\Psi_{\mathrm{G}}}(\hat{x}_{it}^{\mathrm{pg}+}+\hat{x}_{it}^{\mathrm{pg}-})+\\ &\textstyle\sum_{i\in\Psi_{\mathrm{B}}}(\hat{x}_{it}^{\mathrm{ch}+}+\hat{x}_{it}^{\mathrm{ch}-})+\sum_{i\in\Psi_{\mathrm{B}}}(\hat{x}_{it}^{\mathrm{dic}+}+\\ &\textstyle\hat{x}_{it}^{\mathrm{dic}-})+\sum_{i\in\Psi_{\mathrm{R}}}(\hat{x}_{it}^{\mathrm{pr}+}+\hat{x}_{it}^{\mathrm{pr}-})\big]+\\ &\textstyle\sum_{i\in\Psi_{\mathrm{PCC}}}c_i^{\mathrm{ppc,r}}(\hat{x}_{it}^{\mathrm{ppc}+}+\hat{x}_{it}^{\mathrm{ppc}-})\big\}\end{aligned} \tag{A.17}$$

*subject to.*

$$\hat{x}\in\left\{\begin{array}{c}\hat{x}_{it}^{\mathrm{pg}+},\hat{x}_{it}^{\mathrm{pg}-},\hat{x}_{it}^{\mathrm{pr}+},\hat{x}_{it}^{\mathrm{pr}-},\hat{x}_{it}^{\mathrm{ppc}+},\hat{x}_{it}^{\mathrm{ppc}-},\\ \hat{x}_{it}^{\mathrm{ch}+},\hat{x}_{it}^{\mathrm{ch}-},\hat{x}_{it}^{\mathrm{dic}+},\hat{x}_{it}^{\mathrm{dic}-},\hat{x}_{it}^{\mathrm{s}},\hat{x}_{it}^{\mathrm{qg}},\hat{x}_{it}^{\mathrm{qr}},\\ \hat{x}_{it}^{\mathrm{qpc}},x_{ijt}^{\mathrm{p}},x_{ijt}^{\mathrm{q}},x_{ijt}^{\mathrm{l}},x_{it}^{\mathrm{v}}\end{array}\right\} \tag{A.18}$$

$$\begin{aligned}&(x_{it}^{\mathrm{pg}*}+\hat{x}_{it}^{\mathrm{pg}+}-\hat{x}_{it}^{\mathrm{pg}-})+(x_{it}^{\mathrm{pr}*}+\hat{x}_{it}^{\mathrm{pr}+}-\\ &\hat{x}_{it}^{\mathrm{pr}-})+(x_{it}^{\mathrm{dic}*}+\hat{x}_{it}^{\mathrm{dic}+}-\hat{x}_{it}^{\mathrm{dic}-})-(\mathrm{A}.\,x_{it}^{\mathrm{ch}*}+\\ &\hat{x}_{it}^{\mathrm{ch}+}-\hat{x}_{it}^{\mathrm{ch}-})-([\![y_{it}^{\mathrm{ppc}}]\!]^{*}+\hat{x}_{it}^{\mathrm{ppc}+}-\hat{x}_{it}^{\mathrm{ppc}-}+\\ &\textstyle[\![y_{it}^{\mathrm{pf}}]\!])+\sum_{j\in\Omega_i^{+}}(x_{ijt}^{\mathrm{p}}-r_{ij}x_{ijt}^{\mathrm{l}})-\\ &\textstyle\sum_{j\in\Omega_i^{-}}x_{ijt}^{\mathrm{p}}=\check{P}_{it}^{\mathrm{L}},\end{aligned} \tag{A.19a}$$

$$\begin{aligned}&(x_{it}^{\mathrm{qg}*}+\hat{x}_{it}^{\mathrm{qg}})+(x_{it}^{\mathrm{qr}*}+\hat{x}_{it}^{\mathrm{qr}})-([\![y_{it}^{\mathrm{qpc}}]\!]^{*}+\\ &\textstyle\hat{x}_{it}^{\mathrm{qpc}}+[\![y_{it}^{\mathrm{qf}}]\!])+\sum_{j\in\Omega_i^{+}}(x_{ijt}^{\mathrm{q}}-x_{ij}x_{ijt}^{\mathrm{l}})-\\ &\textstyle\sum_{j\in\Omega_i^{-}}x_{ijt}^{\mathrm{q}}=\check{Q}_{it}^{\mathrm{L}},\forall i\in\Psi_{\mathrm{N}},t\in\mathrm{T}_s.\end{aligned} \tag{A.19b}$$

$$x_{it}^{\mathrm{v}}-x_{jt}^{\mathrm{v}}=2(r_{ij}x_{ijt}^{\mathrm{p}}+x_{ij}x_{ijt}^{\mathrm{q}})-(r_{ij}^{2}+x_{ij}^{2})x_{ijt}^{\mathrm{l}}, \tag{A.20}$$

$$\left|x_{ijt}^{\mathrm{l}}+x_{it}^{\mathrm{v}};2\,x_{ijt}^{\mathrm{p}};2\,x_{ijt}^{\mathrm{q}};x_{ijt}^{\mathrm{l}}-x_{it}^{\mathrm{v}}\right|\preceq 0, \tag{A.21}$$

$$0\le x_{ijt}^{\mathrm{l}}\le\overline{\mathrm{L}}_{ij},\forall ij\in\Psi_{\mathrm{f}},t\in\mathrm{T}_s. \tag{A.22}$$

$$\underline{\mathrm{x}}_{it}^{\mathrm{v}}\le x_{it}^{\mathrm{v}}\le\overline{x}_{it}^{\mathrm{v}},\forall i\in\Psi_{\mathrm{N}\setminus\{1\}}, \tag{A.23a}$$

$$x_{1t}^{\mathrm{v}}=1,\forall t\in\mathrm{T}_s. \tag{A.23b}$$

$$0\le\hat{x}_{it}^{\mathrm{pg}+}\le\overline{x}_{it}^{\mathrm{pg}}, \tag{A.24a}$$

$$0\le\hat{x}_{it}^{\mathrm{pg}-}\le\underline{x}_{it}^{\mathrm{pg}}, \tag{A.24b}$$

$$0\le x_{it}^{\mathrm{pg}*}+\hat{x}_{it}^{\mathrm{pg}+}-\hat{x}_{it}^{\mathrm{pg}-}\le\overline{\mathrm{P}}_{it}^{\mathrm{g}}, \tag{A.24c}$$

$$0\le x_{it}^{\mathrm{qg}*}+\hat{x}_{it}^{\mathrm{qg}}\le\overline{\mathrm{Q}}_{i}^{\mathrm{g}},\forall i\in\Psi_{\mathrm{G}},t\in\mathrm{T}_s. \tag{A.24d}$$

$$0\le x_{it}^{\mathrm{pr}*}+\hat{x}_{it}^{\mathrm{pr}+}-\hat{x}_{it}^{\mathrm{pr}-}\le\check{P}_{it}^{\mathrm{g}}, \tag{A.25a}$$

$$0\le x_{it}^{\mathrm{qr}*}+\hat{x}_{it}^{\mathrm{qr}}\le\overline{\mathrm{Q}}_{i}^{\mathrm{gr}},\forall i\in\Psi_{\mathrm{R}},t\in\mathrm{T}_s. \tag{A.25b}$$

$$0\le\hat{x}_{it}^{\mathrm{ch}+}\le\overline{\mathrm{x}}_{it}^{\mathrm{ch}}, \tag{A.26a}$$

$$0\le\hat{x}_{it}^{\mathrm{ch}-}\le\underline{\mathrm{x}}_{it}^{\mathrm{ch}}, \tag{A.26b}$$

$$0\le x_{it}^{\mathrm{ch}*}+\hat{x}_{it}^{\mathrm{ch}+}-\hat{x}_{it}^{\mathrm{ch}-}\le z_{it}^{\mathrm{ch}}\overline{\mathrm{P}}_{i}^{\mathrm{ch}}, \tag{A.26c}$$

$$0\le\hat{x}_{it}^{\mathrm{dic}+}\le\overline{\mathrm{x}}_{it}^{\mathrm{dic}}, \tag{A.27a}$$

$$0\le\hat{x}_{it}^{\mathrm{dic}-}\le\underline{\mathrm{x}}_{it}^{\mathrm{dic}}, \tag{A.27b}$$

$$0\le x_{it}^{\mathrm{dic}*}+\hat{x}_{it}^{\mathrm{dic}+}-\hat{x}_{it}^{\mathrm{dic}-}\le z_{it}^{\mathrm{dic}}\overline{\mathrm{P}}_{i}^{\mathrm{dic}}, \tag{A.27c}$$

$$\begin{aligned}\hat{x}_{i,t+1}^{\mathrm{s}}=\hat{x}_{it}^{\mathrm{s}}+\big[&(x_{it}^{\mathrm{ch}*}+\hat{x}_{it}^{\mathrm{ch}+}-\hat{x}_{it}^{\mathrm{ch}-})\eta_{\mathrm{ch}}-\\ &(x_{it}^{\mathrm{dic}*}+\hat{x}_{it}^{\mathrm{dic}+}-\hat{x}_{it}^{\mathrm{dic}-})/\eta_{\mathrm{dic}}\big]\,\Delta t/\mathrm{E}_i,\end{aligned} \tag{A.28a}$$

$$\underline{\mathrm{s}}\le\hat{x}_{it}^{\mathrm{s}}\le\overline{\mathrm{s}}, \tag{A.28b}$$

$$\hat{x}_{i1}^{\mathrm{s}}=s^{\mathrm{ini}},\hat{x}_{i,\mathrm{T}+1}^{\mathrm{s}}=s^{\mathrm{end}},\forall i\in\Psi_{\mathrm{B}},t\in\mathrm{T}_s. \tag{A.28c}$$

$$\begin{aligned}-\overline{\mathrm{P}}_{i}^{\mathrm{pc}}\le[\![y_{it}^{\mathrm{ppc}}]\!]^{*}+\hat{x}_{it}^{\mathrm{ppc}+}-\hat{x}_{it}^{\mathrm{ppc}-}+[\![y_{it}^{\mathrm{pf}}]\!]\le\\ \overline{\mathrm{P}}_{i}^{\mathrm{pc}},\end{aligned} \tag{A.29a}$$

$$\begin{aligned}-\overline{\mathrm{Q}}_{i}^{\mathrm{pc}}\le[\![y_{it}^{\mathrm{qpc}}]\!]^{*}+\hat{x}_{it}^{\mathrm{qpc}}+[\![y_{it}^{\mathrm{qf}}]\!]\le\overline{\mathrm{Q}}_{i}^{\mathrm{pc}},\\ \forall i\in\Psi_{\mathrm{PCC}},t\in\mathrm{T}_s.\end{aligned} \tag{A.29b}$$

The variable with superscript + denotes the active output of upward flexibility, while that with superscript – is the active output of downward flexibility, and the direction of reactive power adjustment is ignored for brevity. $x_{ijt}^{\mathrm{p}}, x_{ijt}^{\mathrm{q}}, x_{ijt}^{\mathrm{l}}, x_{it}^{\mathrm{v}}$ are system variables of classic SOC-based formulation. Among them, the classic SOC formulation with predetermined power flow direction is adopted in (A.19)-(A.23), where $\check{P}_{it}^{\mathrm{L}}, \check{Q}_{it}^{\mathrm{L}}$ denote possible conditions of nodal demand. (A.24) restricts the upward- and downward- flexibilities provided by upstream power injection, and (A.25) limits regulations of DER under specific estimates $\check{P}_{it}^{\mathrm{g}}$. (A.26)-(A.28) limit flexibility services provided by ESS, and (A.29) restricts the flexibility service exchange at PCC. Notably, DSO will dispatch reserve capacity within a scrolling window $\mathrm{T}_s$ to adapt to rolling forecasts. If it is examined that the fulfillment of flexibility contracts will violate operating constraints, the neutral DSO will indirectly correct the exchange power at PCC by exerting bi-dimensional flexibility services as $\hat{x}_{it}^{\mathrm{ppc}+}, \hat{x}_{it}^{\mathrm{ppc}-}$ in (A.29). It is observed that except for energy trading quantities $[\![y_{it}^{\mathrm{ppc}}]\!]^{*}, [\![y_{it}^{\mathrm{qpc}}]\!]^{*}$ and dispatching plan that have been settled at the first stage, the regulations at PCC submitted by MGs, i.e. $[\![y_{it}^{\mathrm{pf}}]\!]$ and $[\![y_{it}^{\mathrm{qf}}]\!]$, are also fixed at each round in equations (A.19) and (A.29).

## APPENDIX B

## MATHEMATIC FORMULATION OF TWO-STAGE BIDDING MODEL OF MULTIPLE MGS

The mathematical model of the first-stage energy bidding of individual MG is formulated as (B.1)-(B.10).

$$\begin{aligned}\max_{y}\ &\textstyle\sum_{t\in\mathrm{T}}\big\{[\![\tau_{it}^{\mathrm{p}}]\!]*y_{it}^{\mathrm{ppc}}-\\ &\textstyle\big[\sum_{i\in\Psi_{RM}}\mathrm{c}_i^{\mathrm{pr,}}\ \ y_{it}^{\mathrm{pr}}+\sum_{i\in\Psi_{BM}}(\mathrm{c}_i^{\mathrm{ch,M}}y_{it}^{\mathrm{ch}}+\\ &\mathrm{c}_i^{\mathrm{dic,M}}y_{it}^{\mathrm{dic}})+\mathrm{c}_i^{\mathrm{pc,M}}y_{it}^{\mathrm{LpM}}\big]\big\}\end{aligned} \tag{B.1}$$

$$\textit{subject to.}\ y\in\left\{\begin{array}{c}y_{it}^{\mathrm{ppc}},y_{it}^{\mathrm{qpc}},y_{it}^{\mathrm{pr}},y_{it}^{\mathrm{qr}},y_{it}^{\mathrm{ch}},y_{it}^{\mathrm{dic}},y_{it}^{\mathrm{LpM}},\\ y_{it}^{\mathrm{LqM}},y_{it}^{\mathrm{s}},y_{it}^{\mathrm{zch}},y_{it}^{\mathrm{zdic}},\underline{y}_{it}^{\mathrm{ch}},\overline{y}_{it}^{\mathrm{ch}},\underline{y}_{it}^{\mathrm{dic}},\overline{y}_{it}^{\mathrm{dic}}\end{array}\right\} \tag{B.2}$$

$$y_{it}^{\mathrm{ppc}}+y_{it}^{\mathrm{pr}}+y_{it}^{\mathrm{dic}}-y_{it}^{\mathrm{ch}}-\left(\overline{\mathrm{P}}_{it}^{\mathrm{LM}}-y_{it}^{\mathrm{LpM}}\right)=0, \tag{B.3a}$$

$$y_{it}^{\mathrm{qpc}}+y_{it}^{\mathrm{qr}}-y_{it}^{\mathrm{LqM}}=0,\forall i\in\Psi_{\mathrm{PCC}},t\in\mathrm{T}. \tag{B.3b}$$

$$-\overline{\mathrm{P}}_i^{\mathrm{pc}} \leq y_{it}^{\mathrm{ppc}} \leq \overline{\mathrm{P}}_i^{\mathrm{pc}}, \tag{B.4a}$$

$$-\overline{\mathrm{Q}}_i^{\mathrm{pc}} \leq y_{it}^{\mathrm{qpc}} \leq \overline{\mathrm{Q}}_i^{\mathrm{pc}}, \forall i \in \Psi_{\mathrm{PCC}}, t \in \mathrm{T}. \tag{B.4b}$$

$$\mathrm{Prob}\left(0 \leq y_{it}^{\mathrm{pr}} \leq \tilde{P}_{it}^{\mathrm{gM}}\right) \geq 1-\gamma, \tag{B.5a}$$

$$0 \leq y_{it}^{\mathrm{qr}} \leq \overline{\mathrm{Q}}_i^{\mathrm{gM}}, \forall i \in \Psi_{\mathrm{RM}}, t \in \mathrm{T}. \tag{B.5b}$$

$$0 \leq y_{it}^{\mathrm{ch}} - \underline{y}_{it}^{\mathrm{ch}}, \tag{B.6a}$$

$$y_{it}^{\mathrm{ch}} + \overline{y}_{it}^{\mathrm{ch}} \leq y_{it}^{\mathrm{zch}} \overline{\mathrm{P}}_i^{\mathrm{ch}}, \tag{B.6b}$$

$$0 \leq y_{it}^{\mathrm{dic}} - \underline{y}_{it}^{\mathrm{dic}}, \tag{B.7a}$$

$$y_{it}^{\mathrm{dic}} + \overline{y}_{it}^{\mathrm{dic}} \leq y_{it}^{\mathrm{zdic}} \overline{\mathrm{P}}_i^{\mathrm{dic}}, \tag{B.7b}$$

$$y_{it}^{\mathrm{zch}} + y_{it}^{\mathrm{zdic}} \leq 1, y_{it}^{\mathrm{zch}}, y_{it}^{\mathrm{zdic}} \in \{0,1\}, \tag{B.8}$$

$$y_{i,t+1}^{\mathrm{s}} = y_{it}^{\mathrm{s}} + \left(y_{it}^{\mathrm{ch}} \eta_{\mathrm{ch}} - y_{it}^{\mathrm{dic}} / \eta_{\mathrm{dic}}\right) \Delta t / \mathrm{E}_i, \tag{B.9a}$$

$$\underline{\mathrm{s}} \leq y_{it}^{\mathrm{s}} \leq \overline{\mathrm{s}}, \tag{B.9b}$$

$$y_{i1}^{\mathrm{s}} = s^{\mathrm{ini}}, y_{i,\mathrm{T}+1}^{\mathrm{s}} = s^{\mathrm{end}}, \forall i \in \Psi_{\mathrm{BM}}, t \in \mathrm{T}. \tag{B.9c}$$

$$(1-\zeta)\overline{\mathrm{P}}_{it}^{\mathrm{LM}} \leq \overline{\mathrm{P}}_{it}^{\mathrm{LM}} - y_{it}^{\mathrm{LpM}} \leq \overline{\mathrm{P}}_{it}^{\mathrm{LM}}, \tag{B.10a}$$

$$y_{it}^{\mathrm{LqM}} = \left(\overline{\mathrm{P}}_{it}^{\mathrm{LM}} - y_{it}^{\mathrm{LpM}}\right) \tan \theta_{it}, \quad \forall i \in \Psi_{\mathrm{PCC}}, t \in \mathrm{T}. \tag{B.10b}$$

Among them, $y_{it}^{\mathrm{ppc}}, y_{it}^{\mathrm{qpc}}$ are exchanged energy with DSO, $y_{it}^{\mathrm{pr}}, y_{it}^{\mathrm{qr}}$ are active and reactive output of internal DERs, and $y_{it}^{\mathrm{ch}}, y_{it}^{\mathrm{dic}}$ are charging and discharging power of ESS within MG. $y_{it}^{\mathrm{LpM}}, y_{it}^{\mathrm{LqM}}$ denotes the quantity of load shedding. $y_{it}^{\mathrm{s}}$ is the state of charge of ESS, while $y_{it}^{\mathrm{zch}}, y_{it}^{\mathrm{zdic}}$ are binary status variables of ESS. $\underline{y}_{it}^{\mathrm{ch}}, \overline{y}_{it}^{\mathrm{ch}}, \underline{y}_{it}^{\mathrm{dic}}, \overline{y}_{it}^{\mathrm{dic}}$ denote the bi-directional flexibility capacity of ESS within MG. The power balance within nodal MG is restricted by (B.3), and (B.4) limits the quantity of energy exchange at PCC, while (B.6)-(B.7) are operating constraints of ESS which also serves as the only supplier of the reserve within MG. Limited by the capacity of generation resources, the output of DERs is modeled by chance-constraint in (B.5), and the load shedding is permitted in (B.10).

In the second phase, MGs will negotiate flexibility services with others via the P2P sharing in LFM. The model is presented in detail as (B.11)-(B.29).

$$\max_{\hat{y}} \sum_{t \in \mathrm{T}_s} \Big\{ \sum_{j \in \Psi_{\mathrm{PCC}/\{self\}}} \mathrm{c}_i^{\mathrm{sel}} y_{ijt}^{\mathrm{pf}} - \Big[\mathrm{c}_i^{\mathrm{Mr}} \sum_{i \in \Psi_{\mathrm{RM}}} (\hat{y}_{it}^{\mathrm{pr}+} + \hat{y}_{it}^{\mathrm{pr}-}) + \mathrm{c}_i^{\mathrm{Mr}} \sum_{i \in \Psi_{\mathrm{BM}}} (\hat{y}_{it}^{\mathrm{ch}+} + \hat{y}_{it}^{\mathrm{ch}-} + \hat{y}_{it}^{\mathrm{dic}+} + \hat{y}_{it}^{\mathrm{dic}-}) + \mathrm{c}_i^{\mathrm{pc},} \ (\hat{y}_{it}^{\mathrm{LpM}} - \hat{y}_{it}^{\mathrm{Lpg}}) + \sum_{k \in \Psi_{\mathrm{PCC}/\{self\}}} \mathrm{c}_k^{\mathrm{pur}} y_{kit}^{\mathrm{pf}} + \mathrm{c}_i^{\mathrm{ppc,Mr}} (\hat{y}_{it}^{\mathrm{ppc}+} + \hat{y}_{it}^{\mathrm{ppc}-}) \Big] \Big\} \tag{B.11}$$

*subject to.*

$$\hat{y} \in \left\{ \begin{matrix} \hat{y}_{ijt}^{\mathrm{pf}}, \hat{y}_{ijt}^{\mathrm{qf}}, \hat{y}_{it}^{\mathrm{ppc}+}, \hat{y}_{it}^{\mathrm{ppc}-}, \hat{y}_{it}^{\mathrm{pr}+}, \hat{y}_{it}^{\mathrm{pr}-}, \hat{y}_{it}^{\mathrm{ch}+}, \\ \hat{y}_{it}^{\mathrm{ch}-}, \hat{y}_{it}^{\mathrm{dic}+}, \hat{y}_{it}^{\mathrm{dic}-}, \hat{y}_{it}^{\mathrm{s}}, \hat{y}_{it}^{\mathrm{qr}}, \hat{y}_{it}^{\mathrm{qpc}}, \hat{y}_{it}^{\mathrm{Lpg}}, \hat{y}_{it}^{\mathrm{LpM}} \end{matrix} \right\} \tag{B.12}$$

$$[\![ y_{it}^{\mathrm{ppc}*} ]\!] + \sum_{k \in \Psi_{\mathrm{PCC}/\{self\}}} \hat{y}_{kit}^{\mathrm{pf}} - \sum_{j \in \Psi_{\mathrm{PCC}/\{self\}}} \hat{y}_{ijt}^{\mathrm{pf}} + [\![ \hat{x}_{it}^{\mathrm{ppc}+} ]\!] - [\![ \hat{x}_{it}^{\mathrm{ppc}-} ]\!] + \hat{y}_{it}^{\mathrm{ppc}+} - \hat{y}_{it}^{\mathrm{ppc}-} + (y_{it}^{\mathrm{pr}*} + \hat{y}_{it}^{\mathrm{pr}+} - \hat{y}_{it}^{\mathrm{pr}-}) + (y_{it}^{\mathrm{dic}*} + \hat{y}_{it}^{\mathrm{dic}+} - \hat{y}_{it}^{\mathrm{dic}-}) - (y_{it}^{\mathrm{ch}*} + \hat{y}_{it}^{\mathrm{ch}+} - \hat{y}_{it}^{\mathrm{ch}-}) - (y_{it}^{\mathrm{LpM}*} + \hat{y}_{it}^{\mathrm{Lpg}} - \hat{y}_{it}^{\mathrm{LpM}}) = 0, \tag{B.13}$$

$$[\![ y_{it}^{\mathrm{qpc}*} ]\!] + \sum_{k \in \Psi_{\frac{\mathrm{PCC}}{\{self\}}}} \hat{y}_{kit}^{\mathrm{qf}} - \sum_{j \in \Psi_{\frac{\mathrm{PCC}}{\{self\}}}} \hat{y}_{ijt}^{\mathrm{qf}} + [\![ \hat{x}_{it}^{\mathrm{qpc}} ]\!] + \hat{y}_{it}^{\mathrm{qpc}} = \check{Q}_{it}^{\mathrm{LM}} - (y_{it}^{\mathrm{LqM}} + \hat{y}_{it}^{\mathrm{Lqg}}), \quad \forall i \in \Psi_{\mathrm{PCC}}, t \in \mathrm{T}_s. \tag{B.14}$$

$$-\overline{\mathrm{P}}_i^{\mathrm{pc}} \leq [\![ y_{it}^{\mathrm{ppc}*} ]\!] + \sum_k \hat{y}_{kit}^{\mathrm{pf}} - \sum_j \hat{y}_{ijt}^{\mathrm{pf}} + [\![ \hat{x}_{it}^{\mathrm{ppc}+} ]\!] - [\![ \hat{x}_{it}^{\mathrm{ppc}-} ]\!] + \hat{y}_{it}^{\mathrm{ppc}+} - \hat{y}_{it}^{\mathrm{ppc}-} \leq \overline{\mathrm{P}}_i^{\mathrm{pc}}, \tag{B.15}$$

$$-\overline{\mathrm{Q}}_i^{\mathrm{pc}} \leq [\![ y_{it}^{\mathrm{qpc}*} ]\!] + \sum_{k \in \Psi_{\frac{\mathrm{PCC}}{\{self\}}}} \hat{y}_{kit}^{\mathrm{qf}} - \sum_{j \in \Psi_{\frac{\mathrm{PCC}}{\{self\}}}} \hat{y}_{ijt}^{\mathrm{qf}} + [\![ \hat{x}_{it}^{\mathrm{qpc}} ]\!] + \hat{y}_{it}^{\mathrm{qpc}} \leq \overline{\mathrm{Q}}_i^{\mathrm{pc}}, \quad \forall i \in \Psi_{\mathrm{PCC}}, t \in \mathrm{T}_s. \tag{B.16}$$

$$0 \leq y_{it}^{\mathrm{pr}*} + \hat{y}_{it}^{\mathrm{pr}+} - \hat{y}_{it}^{\mathrm{pr}-} \leq \check{P}_{it}^{\mathrm{gM}}, \tag{B.17}$$

$$0 \leq y_{it}^{\mathrm{qr}*} + \hat{y}_{it}^{\mathrm{qr}} \leq \overline{\mathrm{Q}}_i^{\mathrm{gM}}, \forall i \in \Psi_{\mathrm{RM}}, t \in \mathrm{T}_s. \tag{B.18}$$

$$0 \leq \hat{y}_{it}^{\mathrm{ch}+} \leq \overline{y}_{it}^{\mathrm{ch}}, \tag{B.19}$$

$$0 \leq \hat{y}_{it}^{\mathrm{ch}-} \leq \underline{y}_{it}^{\mathrm{ch}}, \tag{B.20}$$

$$0 \leq y_{it}^{\mathrm{ch}*} + \hat{y}_{it}^{\mathrm{ch}+} - \hat{y}_{it}^{\mathrm{ch}-} \leq y_{it}^{\mathrm{zch}*} \overline{\mathrm{P}}_i^{\mathrm{ch,M}}, \tag{B.21}$$

$$0 \leq \hat{y}_{it}^{\mathrm{dic}+} \leq \overline{y}_{it}^{\mathrm{dic}}, \tag{B.22}$$

$$0 \leq \hat{y}_{it}^{\mathrm{dic}-} \leq \underline{y}_{it}^{\mathrm{dic}}, \tag{B.23}$$

$$0 \leq y_{it}^{\mathrm{dic}*} + \hat{y}_{it}^{\mathrm{dic}+} - \hat{y}_{it}^{\mathrm{dic}-} \leq y_{it}^{\mathrm{zdic}*} \overline{\mathrm{P}}_i^{\mathrm{dic},M}, \tag{B.24}$$

$$\hat{y}_{i,t+1}^{\mathrm{s}} = \hat{y}_{it}^{\mathrm{s}} + \left[(y_{it}^{\mathrm{ch}*} + \hat{y}_{it}^{\mathrm{ch}+} - \hat{y}_{it}^{\mathrm{ch}-}) \eta_{\mathrm{ch}} - (y_{it}^{\mathrm{dic}*} + \hat{y}_{it}^{\mathrm{dic}+} - \hat{y}_{it}^{\mathrm{dic}-}) / \eta_{\mathrm{dic}}\right] \Delta t / \mathrm{E}_i^{\mathrm{M}}, \tag{B.25}$$

$$\underline{\mathrm{s}} \leq \hat{y}_{it}^{\mathrm{s}} \leq \overline{\mathrm{s}}, \tag{B.26}$$

$$\hat{y}_{i1}^{\mathrm{s}} = s^{\mathrm{ini}}, \hat{y}_{i,\mathrm{T}+1}^{\mathrm{s}} = s^{\mathrm{end}}, \forall i \in \Psi_{\mathrm{BM}}, t \in \mathrm{T}_s. \tag{B.27}$$

$$(1-\zeta)\check{P}_{it}^{\mathrm{LM}} \leq \check{P}_{it}^{\mathrm{LM}} - \left(y_{it}^{\mathrm{LpM}*} + \hat{y}_{it}^{\mathrm{Lpg}} - \hat{y}_{it}^{\mathrm{LpM}}\right) \leq \check{P}_{it}^{\mathrm{LM}}, \tag{B.28}$$

$$\hat{y}_{it}^{\mathrm{LqM}} = \left(\hat{y}_{it}^{\mathrm{Lpg}} - \hat{y}_{it}^{\mathrm{LpM}}\right) \tan \theta_{i,t}, \quad \forall i \in \Psi_{\mathrm{PCC}}, t \in \mathrm{T}_s. \tag{B.29}$$

Where $\hat{y}_{ijt}^{\mathrm{pf}}, \hat{y}_{ijt}^{\mathrm{qf}}$ denote the active and reactive flexibility services of MG j purchased via P2P trading from MG I, and the variable of superscript +,- are upward and downward flexibility services, and others are adjustment quantities of corresponding variables. Among them, (B.13)-(B.14) restrict the nodal power balance, and (B.15)-(B.16) are operating constraints of energy exchange at PCC. (B.17)-(B.18) limit the available output of DERs with updated $\check{P}_{i,t}^{\mathrm{gM}}$. (B.19)-(B.27) describe the operating conditions of internal ESS when supplying reserves, and (B.28)-(B.29) restricts the altered load control in the face of demand scenarios. It is worth noting that DSO regulates flexibility contracts by indirectly introducing bi-directional adjustments in (B.13)-(B.16), i.e. $[\![ \hat{x}_{i,t}^{\mathrm{ppc}+} ]\!], [\![ \hat{x}_{i,t}^{\mathrm{ppc}-} ]\!], [\![ \hat{x}_{i,t}^{\mathrm{qpc}} ]\!]$, while the window value transferred by MG to DSO is the summation of the second-stage decision of MG, i.e. $\hat{y}_{i,j,t}^{\mathrm{pf}} = \sum_k \hat{y}_{k,i,t}^{\mathrm{pf}} - \sum_j \hat{y}_{i,j,t}^{\mathrm{pf}} + \hat{y}_{i,t}^{\mathrm{ppc}+} - \hat{y}_{i,t}^{\mathrm{ppc}-}$.

## APPENDIX C
## MATHEMATIC FORMULATIONS OF CC-UDLMP

To remove calculation obstacles, we performed recursive calculations to obtain this price by fixing the power flow path in the convex dual counterpart, formulated as (C.1)-(C.16).

$$\begin{aligned}&\max \textstyle\sum_{t\in \mathrm{T}} \Big[\sum_{i\in\Psi_N}\big(\tilde{P}^{\mathrm{L}}_{it}\tau^{\mathrm{p}}_{it} + \tilde{Q}^{\mathrm{L}}_{it}\tau^{\mathrm{q}}_{it}\big) + \\ &\textstyle\sum_{i\in\Psi_{\mathrm{R}}}\big(\bar{P}^{*}_{u}\varphi^{\mathrm{pr}}_{it} + \bar{\mathrm{Q}}^{\mathrm{g}}_{it}\varphi^{\mathrm{qr}}_{it}\big) + \sum_{i\in\Psi_{N\backslash\{1\}}}\big(\underline{v}^{*}_{it}\lambda^{v}_{it} + \\ &\overline{v}^{*}_{it}\varphi^{v}_{it}\big) + \omega^{\mathrm{bn}}_{t} + \textstyle\sum_{ij\in\Psi_f}\varphi^{z+}_{it}\bar{\mathrm{L}}^{*}_{ij} + \\ &\textstyle\sum_{i\in\Psi_{\mathrm{G}}}\Big(\overline{P}^{\mathrm{g}}_{i,t}\varphi^{\mathrm{pg}}_{it} + \overline{Q}^{g}_{i,t}\varphi^{\mathrm{qg}}_{it}\Big) + \sum_{i\in\Psi_B}\big(\bar{x}^{\mathrm{s}}\lambda^{\mathrm{s}}_{i,t} + \\ &\underline{x}^{\mathrm{s}}\varphi^{\mathrm{s}}_{i,t}\big)\Big] + \textstyle\sum_{i\in\Psi_{\mathrm{B}}}\big(\bar{x}^{\mathrm{ini}}\omega^{\mathrm{ini}}_{i} + \bar{x}^{\mathrm{end}}\omega^{\mathrm{end}}_{i}\big)\end{aligned} \quad \text{(C.1)}$$

$$\textit{subject to.}\quad x^{\mathrm{pg}}_{i,t}\colon \tau^{\mathrm{p}}_{it} + \lambda^{\mathrm{pg}}_{it} + \varphi^{\mathrm{pg}}_{it} = \mathrm{c}^{\mathrm{pg}}_{i}, \quad \text{(C.2)}$$

$$x^{\mathrm{qg}}_{i,t}\colon \tau^{\mathrm{q}}_{it} + \lambda^{\mathrm{qg}}_{it} + \varphi^{\mathrm{qg}}_{it} = 0, \forall i \in \Psi_{\mathrm{G}}. \quad \text{(C.3)}$$

$$x^{\mathrm{pr}}_{i,t}\colon \tau^{\mathrm{p}}_{it} + \lambda^{\mathrm{pr}}_{it} + \varphi^{\mathrm{pr}}_{it} = \mathrm{c}^{\mathrm{pr}}_{i}, \quad \text{(C.4)}$$

$$x^{\mathrm{qr}}_{i,t}\colon \tau^{\mathrm{q}}_{it} + \lambda^{\mathrm{qr}}_{it} + \varphi^{\mathrm{qr}}_{it} = 0, \forall i \in \Psi_{\mathrm{R}}. \quad \text{(C.5)}$$

$$\begin{aligned}\tilde{x}^{v}_{1,t}\colon &\textstyle\sum_{j\in\Omega^{+}_{i}}\big(\omega_{ijt} + d^{+}_{ijt} - d^{\mathrm{C}+}_{ijt}\big) - \sum_{k\in\Omega^{-}_{i}}\omega_{kit} + \\ &\omega^{\mathrm{bn}}_{t} = 0, \forall i \in \Psi_{\{1\}},\end{aligned} \quad \text{(C.6)}$$

$$\begin{aligned}\tilde{x}^{v}_{i,t}\colon &\textstyle\sum_{j\in\Omega^{+}_{i}}\big(\omega_{ijt} + d^{+}_{ijt} - d^{\mathrm{C}+}_{ijt}\big) - \sum_{k\in\Omega^{-}_{i}}\omega_{kit} + \\ &\lambda^{v}_{it} + \varphi^{v}_{it} = 0, \forall i \in \Psi_{\mathrm{N}\backslash\{1\}}, t \in \mathrm{T}.\end{aligned} \quad \text{(C.7)}$$

$$\begin{aligned}x^{\mathrm{p}+}_{ij,t}\colon &-\tau^{\mathrm{p}}_{it} + \tau^{\mathrm{p}}_{jt} - 2r_{ij}\omega_{ijt} + d^{\mathrm{A}+}_{ijt} + \lambda^{\mathrm{p}+}_{ijt} + \\ &\varphi^{\mathrm{p}+}_{ijt} = 0,\end{aligned} \quad \text{(C.8)}$$

$$\begin{aligned}x^{\mathrm{q}+}_{ij,t}\colon &-\tau^{\mathrm{q}}_{it} + \tau^{\mathrm{q}}_{jt} - 2x_{ij}\omega_{ijt} + d^{\mathrm{B}+}_{ijt} + \lambda^{\mathrm{q}+}_{ijt} + \\ &\varphi^{\mathrm{q}+}_{ijt} = 0,\end{aligned} \quad \text{(C.9)}$$

$$\begin{aligned}\tilde{x}^{l+}_{ij,t}\colon &(r^{2}_{ij} + x^{2}_{ij})\omega_{ijt} + \lambda^{f+}_{ijt} + \varphi^{f+}_{ijt} + d^{+}_{ijt} + d^{\mathrm{C}+}_{ijt} \\ &- r_{ij}\tau^{\mathrm{p}}_{it} - x_{ij}\tau^{\mathrm{q}}_{it} = \mathrm{c}^{\mathrm{nl}}r_{ij}, \\ &\forall ij \in \Psi_{\mathrm{f}}, t \in \mathrm{T}.\end{aligned} \quad \text{(C.10)}$$

$$\big|d^{+}_{ijt}\,; d^{\mathrm{A}+}_{ijt}\,; d^{\mathrm{B}+}_{ijt}; d^{\mathrm{C}+}_{ijt}\big| \preccurlyeq 0, \quad \text{(C.11)}$$

$$x^{\mathrm{ch}}_{i,t}\colon -\tau^{\mathrm{p}}_{it} - \omega^{\mathrm{s}}_{i,t}\eta_{\mathrm{ch}}\Delta t/\mathrm{E}_{i} + \lambda^{\mathrm{ch}}_{it} + \varphi^{\mathrm{ch}}_{it} = \mathrm{c}^{\mathrm{ch}}_{i}, \quad \text{(C.12)}$$

$$\begin{aligned}x^{\mathrm{dic}}_{i,t}\colon &\tau^{\mathrm{p}}_{it} + \omega^{\mathrm{s}}_{it}\Delta t/(\eta_{\mathrm{dic}}\mathrm{E}_{i}) + \lambda^{\mathrm{dic}}_{it} + \varphi^{\mathrm{dic}}_{it} = \mathrm{c}^{\mathrm{dic}}_{i}, \\ &\forall i \in \Psi_{\mathrm{B}}, t \in \mathrm{T}.\end{aligned} \quad \text{(C.13)}$$

$$x^{\mathrm{s}}_{i,1}\colon -\omega^{\mathrm{s}}_{i1} + \omega^{\mathrm{ini}}_{i} = 0, \forall i \in \Psi_{\mathrm{B}}. \quad \text{(C.14)}$$

$$\begin{aligned}x^{\mathrm{s}}_{i,t}\colon &\omega^{\mathrm{s}}_{it-1} - \omega^{\mathrm{s}}_{it} + \lambda^{\mathrm{s}}_{it} + \varphi^{\mathrm{s}}_{it} = 0, \\ &\forall i \in \Psi_{\mathrm{B}}, t \in \mathrm{T}.\end{aligned} \quad \text{(C.15)}$$

$$x^{\mathrm{s}}_{i,\mathrm{T}+1}\colon \omega^{\mathrm{s}}_{iT} + \omega^{\mathrm{end}}_{i} = 0, \forall i \in \Psi_{\mathrm{B}}. \quad \text{(C.16)}$$

In contrast to the application of Karush-Kuhn-Tucker (KKT) conditions in [30], the recursive process does not involve complementary conditions, avoiding computational expenses with the participation of conditional variables. Moreover, it is verified that compared with the duality relaxation which simplifies complementarity by minimizing the duality gap in [13], the proposed formation possesses superior performance with acceptable calculation accuracy.

## APPENDIX D

### EXISTENCE OF NASH EQUILIBRIUM IN FULLY DISTRIBUTED ITERATIVE ALGORITHM

From the perspective of the decision-making priority, DSO and multiple MGs can be regarded as participants at different layers, and their action update is exclusively triggered by new knowledge announced by another.

In the first stage, individual MG submits energy-related bids to DSO simultaneously. This interactive course can be regarded as a recursive decision on the CC-UDLMP at PCC. By substituting (46.b) into (46.a), the formulation of CC-UDLMP at the *(k+1)*-th iteration is rewritten as (D.1). $\widehat{\mathcal{R}}$ represents the combined reaction function of DSO after collecting all bids from MGs, and $\widehat{\mathcal{R}}_c$ is the nested calculation as $\widehat{\mathcal{R}}\big(\mathcal{R}_b(\cdot)\big)$.

$$\begin{aligned}\tau^{\mathrm{p}}_{k+1} = \mathcal{R}^{+}\big(y_k, \mathcal{R}(y_k)\big) = \widehat{\mathcal{R}}(y_k) = \widehat{\mathcal{R}}\Big(\mathcal{R}_b(\tau^{p}_{k})\Big) \\ = \widehat{\mathcal{R}}_c(\tau^{\mathrm{p}}_{k})\end{aligned} \quad \text{(D.1)}$$

It is observed that the formation of CC-UDLMP at the next round is closely correlated with the result obtained from the previous round. Assumed that augmented Lagrange functions of all reaction models are continuously differentiable and optimal solutions are regular points without causing inequality degeneration [31]-[32], then, the sensitivity matrix derived from the second-order optimality condition will be valid in the neighborhoods of the optimal point when perturbing or modifying parameters [31]. Based on the Banach contraction mapping theorem, this calculation will converge to a fixed point only when the recursive function $\widehat{\mathcal{R}}_c$ satisfies the contraction property. By applying chain rules in the above reaction models, it is verified that the spectral radius of $\widehat{\mathcal{R}}_c$ is significantly less than 1, thereby verifying the convergence of iterative CC-UDLMP formation during the iterative algorithm.

Analogously, the reactions of DSO and multiple MGs in (47) are formed as (D.2) and (D.3) respectively. Notably, as these composite functions also satisfy contraction conditions, their actions will be convergent as well.

$$\textbf{DSO (1}^{\textbf{st}}\textbf{ stage):}\ (x_{k+1}, z_{k+1}) = \mathcal{R}\Big(\mathcal{R}_b(\tau^{\mathrm{p}}_{k})\Big) \quad \text{(D.2)}$$

$$\textbf{Microgrids (1}^{\textbf{st}}\textbf{ stage):}\ \ y_{k+1} = \mathcal{R}_b\big(\tau^{\mathrm{p}}_{k+1}\big) \quad \text{(D.3)}$$

As for the second-stage flexibility trading, considering the strategic bidding of MG is independent, and the pivotal principle of flexibility trading between MGs is to seek a win-to-win solution, this decision-making process of MGs can be regarded as a cooperative game. In other words, a MG coalition is formed to generate P2P flexibility contracts within the centralized model. By substituting (47.b) in (47.a) and simplifying the fixed first-stage result in expressions, the reaction of MG coalition at the *(l+1)*-round is formed as (D.4).

$$\textbf{Microgrid Coalition:}\ \ \hat{y}_{l+1} = \acute{\mathcal{R}}_b(\acute{\mathcal{R}}(\hat{y}_l)) \quad \text{(D.4)}$$

It is easy to examine that the compound reaction function $\acute{\mathcal{R}}_b(\acute{\mathcal{R}}(\cdot))$ fulfills the contraction property, illustrating that the recursive calculation of the movement of the MG coalition will converge to a stationary point. Analogously, the reaction of DSO at the same round in (47.b) is reformulated as (D.5). It is noted that as the spectral radius of the function $\acute{\mathcal{R}}$ is less than 1, the newly nested multiplication will not affect the contraction property of (D.5), reflecting the convergence of this decision.

$$\textbf{DSO (2}^{\textbf{nd}}\textbf{ stage):}\ \ \hat{x}_{l+1} = \acute{\mathfrak{R}}(\hat{y}_{l+1}) \quad \text{(D.5)}$$